\documentclass[]{aa}
\usepackage{graphicx}
\usepackage{natbib}
\bibpunct{(}{)}{;}{a}{}{,}
\graphicspath{{figures/}}
\usepackage{hyperref}
\usepackage{algorithm,algorithmicx}

\begin{document}

%
%
\title{A new approach for modelling chromospheric evaporation in response to enhanced coronal heating: 1 the method}
\author{
C. D. Johnston\inst{1} \and A. W. Hood\inst{1} 
\and P. J. Cargill\inst{1, 2}
\and I. De Moortel\inst{1} 
}
\institute{School of Mathematics and Statistics, University of St. Andrews, St. Andrews, Fife, KY16 9SS, UK.
\and
Space and Atmospheric Physics, The Blackett Laboratory, Imperial College, London, SW7 2BW, UK.
\\
\email{cdj3@st-andrews.ac.uk}
}

%
%

  \abstract
  {
  We present a new computational approach that addresses the 
  difficulty of obtaining the correct interaction between the 
  solar corona and the transition region in response to rapid 
  heating events. In the coupled corona, transition 
  region and chromosphere system, an enhanced downward 
  conductive 
  flux results in an upflow (chromospheric evaporation). 
  However, obtaining the correct upflow generally requires 
  high spatial resolution in order to resolve the 
  transition region. With an unresolved transition region, 
  artificially low coronal densities are obtained
  because the
  downward heat flux \lq jumps\rq\ across the unresolved 
  region to the
  chromosphere, underestimating the upflows.
  Here, we treat the lower transition 
  region as a discontinuity that responds to changing coronal 
  conditions through the imposition of a jump condition
  that is 
  derived from an integrated form of energy conservation. To 
  illustrate and benchmark this approach
  against a fully resolved one-dimensional model, we present 
  field-aligned 
  simulations of coronal loops in response to a range 
  of impulsive (spatially uniform) heating events. We show 
  that our approach leads to a significant improvement in the 
  coronal density evolution than just when using coarse 
  spatial 
  resolutions insufficient to resolve the lower transition 
  region. Our approach compensates for the jumping of 
  the heat flux by imposing a velocity correction that ensures 
  that the energy from the
  heat flux goes into driving the transition region 
  dynamics, rather than being lost through radiation. Hence, 
  it is possible to obtain improved coronal densities. 
  The advantages of 
  using this approach in both one-dimensional hydrodynamic and 
  three-dimensional magnetohydrodynamic simulations are 
  discussed.
  }
  \titlerunning{A new approach for modelling chromospheric
  evaporation}
  \maketitle

%
%

\section{Introduction
  \label{section:Introduction}}
  \indent
  The interaction between the solar corona and chromosphere is 
  central to understanding the observed properties of 
  magnetically closed coronal loops. It is well known that if 
  the corona is heated impulsively (by for example, a flare, 
  microflare or 
  nanoflare), both the temperature and density increase and 
  then decline, with the time of peak temperature preceding 
  that of the peak density. The changes in density can only be 
  accounted for by mass exchange between the corona and 
  chromosphere, mediated by the transition region (TR). 
  \\
  \indent
  Recognising the role of the TR is essential for developing
  reliable models of impulsive heating. For a static 
  equilibrium loop with steady heating, the TR is defined as 
  the region extending from the top of the chromosphere to the 
  location where thermal conduction changes from an energy 
  loss to a gain \citep[e.g.][]{paper:Veseckyetal1979}. The 
  full TR 
  occupies 
  roughly 10\% of the total loop length, the radiation from it 
  is roughly twice that from the corona, and the temperature 
  at its top is of order 60\% the temperature at the loop apex 
  \citep{paper:Cargilletal2012a}. The energy balance in the TR 
  is 
  approximately between downward thermal conduction and 
  optically 
  thin radiation (for a loop in thermal equilibrium).
  \\
  \indent 
  The change in coronal density in response to impulsive 
  heating arises because the increased coronal temperature 
  implied by the heating gives rise to an excess downward heat 
  flux that the TR is unable to radiate 
  \citep{paper:Klimchuketal2008,paper:Cargilletal2012a}.
  The outcome is an enthalpy flux 
  from chromosphere, through the TR, to the corona, often 
  called (chromospheric) \lq evaporation\rq\
  \citep[e.g.][]{paper:Antiochos&Sturrock1978}. 
  The location of the TR moves downward in 
  the atmosphere, and the evaporation process actually heats 
  chromospheric material to coronal temperatures. The 
  process is reversed after the density peaks when the TR 
  requires a larger heat flux than the corona can provide, and 
  so instead an enthalpy flux from the corona is set up, which 
  both drains the corona and powers the TR radiative losses 
  \citep{paper:Bradshaw&Cargill2010a,
  paper:Bradshaw&Cargill2010b}. The TR now moves upwards as 
  the chromosphere 
  is replenished.
  \\
  \indent
  While straightforward in principle, this heating and upflow
  followed by
  cooling and downflow cycle poses major challenges for 
  computational modelling, with conductive cooling being the 
  most severe. For a loop in static equilibrium, in the TR one 
  has an approximate energy 
  equation that equates,
  \begin{align}
    \hspace{2cm}
    \kappa_0 T^{7/2}/ L_T^2 \sim (P/2k_B)^2 \Lambda(T)/T^2,
  \end{align}
  where $L_T$ is the temperature length scale (see Eq. 
  \eqref{eqn:1d_L_T}
  for the definition)
  and the radiative loss 
  function $\Lambda(T)$ decreases as a function of temperature 
  above 
  $10^5$K. Thus, one finds $L_T^2 \sim T^{11/2} / \Lambda(T)$, 
  assuming 
  the 
  pressure is constant. Since $T$ decreases in the TR, $L_T$ 
  must 
  also decrease rapidly. For a static loop with peak 
  temperature 1.75MK and density 0.25$\times 10^{15}$m$^{-3}$,
  $L_T \sim 30$km at 
  $10^5$K. 
  When impulsive heating occurs, $L_T$ is even smaller. This 
  leads to the familiar difficulty with computational models 
  of loop evolution: how to implement a grid that resolves 
  the TR. Good resolution is essential in order to obtain the 
  correct coronal density
  \citep[][hereafter BC13]{paper:Bradshaw&Cargill2013}, 
  otherwise the downward 
  heat flux jumps over an under-resolved TR to the 
  chromosphere where the energy is radiated away. 
  BC13 showed 
  that major errors in the coronal density were likely with 
  lack of resolution.
  \\
  \indent
  Since the conductive timescale across a grid point
   has real physical meaning for 
  the problems at hand, an explicit numerical method is to be 
  preferred (implicit solvers require matrix inversion with no 
  guarantee of convergence). One option is to 
  use brute force on a fixed grid with a large number of grid 
  points. This is slow, since numerical stability of an 
  explicit algorithm requires
  $
  \Delta t \leq\textrm{min}(
  k_B n (\Delta z)^2 
  /
  (2\kappa_0
  T^{5/2})
  )
  $
  (where $\Delta z$ is the 
  cell 
  width
  and the timestep is the minimum over the whole grid), so 
  that 
  a lot of time is wasted computing in the corona where $L_T$ 
  is 
  large and high spatial resolution is not required. A 
  non-uniform fixed grid, with points localised at the TR is 
  an 
  option, 
  but since the TR moves (see above), there is no guarantee 
  that high resolution will be where it is required. Instead, 
  modern schemes use an adaptive mesh which allocates points 
  where they are needed 
  \citep[][BC13]{paper:Bettaetal1997,
  paper:Bradshaw&Mason2003}. The 
  time step restriction is the same as for a uniform grid, but 
  effort is no 
  longer wasted computing highly resolved coronal solutions.
  \\
  \indent 
  Thus far we have not distinguished between the common 
  one-dimensional (1D) 
  hydrodynamic (field-aligned) modelling and 
  multi-dimensional MHD simulations. It is straightforward for 
  a 1D 
  code with an adaptive mesh and a large computer to model a 
  single heating event, and, with patience, to model a 
  nanoflare train lasting several tens of thousands of seconds 
  \citep{paper:Cargilletal2015}. 
  However, ensembles of thousands of 
  loop strands heated by nanoflares pose more severe 
  computational challenges. This has led to the development of 
  zero-dimensional field-aligned hydrodynamical models 
  \citep[e.g.][]
  {paper:Klimchuketal2008,paper:Cargilletal2012a,
  paper:Cargilletal2012b,paper:Cargilletal2015}
  that provide a 
  quick and accurate answer to the coronal 
  response of a 
  loop to heating. 
  \\
  \indent
  The implementation of field-aligned loop plasma evolution 
  into multi-dimensional MHD models poses much more serious 
  challenges due to the number of grid 
  points that can be used, so that
  3D MHD 
  simulations run in a realistic time. This is of the order
  of $500^3$ 
  at the present time.
  If one 
  desires to resolve 
  the TR with 
  a fixed 
  grid, 
  one needs several thousand points in one direction, so that 
  there will be a loss of resolution elsewhere 
  as well as a potentially crippling reduction of the time 
  step. 
  \\
  \indent
  The second difficulty is that while an adaptive mesh can 
  still be used in the TR, with commensurate computational 
  benefits, there can be other parts of such simulations that 
  have equally pressing requirements for high resolution, such 
  as current sheets, and, once 
  again, an adaptive mesh does not eliminate the time step 
  problem.
  \\
  \indent
  Artificially low coronal densities is the
  main consequence of not   
  resolving the TR
  (BC13) and this has significant
  implications for coronal modelling. 
  The purpose of this paper is to present
  a physically motivated approach to 
  deal with this problem 
  by using an integrated form of energy 
  conservation that treats the unresolved region of the 
  lower TR (referred
  to as the unresolved transition region) as a 
  discontinuity, that responds to changing
  coronal conditions through the imposition of a 
  jump condition.
  \\
  \indent
  We describe the key features of the 1D field-aligned model
  and the definitions used to locate the unresolved transition
  region (UTR) 
  in Section \ref{section:Equations and Numerical Method}
  and Appendix \ref{app:A}. 
  The UTR jump condition is  
  derived and the implementation described in Section
  \ref{section:Unresolved Transition Region Jump Condition}. 
  In Section \ref{section:Results}, we present 
  example simulations to benchmark our approach against a
  fully resolved 1D model.
  We conclude with a discussion of our new approach and the
  advantages of employing it, in both 1D and 3D
  simulations, in Section 
  \ref{section:Discussion and Conclusions}.
  
%
%

\section{Equations and numerical method
  \label{section:Equations and Numerical Method}}
  \indent
  In this work we model chromospheric evaporation in response 
  to enhanced impulsive coronal heating by considering the 1D 
  field-aligned MHD
  equations for a single magnetic strand, with uniform 
  cross-section,
  \begin{align}
    & 
    \frac{\partial\rho}{\partial t} + v\frac{\partial\rho}
    {\partial z} = - \rho\frac{\partial v}{\partial z}, 
    \label{eqn:1d_continuity}
    \\[2mm]
    & 
    \rho \frac{\partial v}{\partial t} + \rho v
    \frac{\partial 
    v}
    {\partial z}
    = -
    \frac{\partial P}
    {\partial z} - \rho g_{\parallel} 
    + \rho \nu \frac{\partial^2 v}{\partial z^2},
    \label{eqn:1d_motion}
    \\[2mm]
    & 
    \rho \frac{\partial\epsilon}{\partial t} + \rho v
    \frac{\partial
    \epsilon}
    {\partial z}
    = -P\frac{\partial 
    v}
    {\partial z} - 
    	\frac{\partial F_c}{\partial z} + Q
    - \! n^2 \Lambda(T)
    \!
    + 
    \!
    \rho \nu 
    \!   
    \left(
    \frac{\partial v}{\partial z}
    \right)^{\! \! \!2} \! , \!  \! 
    \label{eqn:1d_tee}
    \\[2mm]
    & 
    P = 2 \, k_B n T.
    \label{eqn:gas_law}
  \end{align} 
  Here, $z$ is the spatial coordinate along the magnetic 
  field, 
  $\rho$ is the mass density, $P$ is the gas pressure, $T$ is 
  the temperature,  $k_B$ is the Boltzmann 
  constant, 
  $\epsilon=P/(\gamma-1) \rho $ is the specific 
  internal energy density,
  $n$ is the number density ($n=\rho/1.2m_p$, $m_p$ is the
  proton mass),
  $v$ is the velocity parallel to the 
  magnetic field, $g_{\parallel}$ is the field-aligned 
  gravitational acceleration (for which we use a profile 
  that corresponds to a semi-circular 
  strand),
  $\nu$ is the viscosity (shock viscosity is also included
  as discussed in \cite{paper:Arber2001}),
  $F_c=-\kappa_0 T^{5/2}\partial T/\partial z$
  is the heat flux,
  $Q$ is the volumetric heating rate and $\Lambda(T)$ is the 
  optically thin radiative loss function for which we use the 
  piecewise continuous form defined in
  \cite{paper:Klimchuketal2008}.
  \\
  \indent
  We solve the 1D field-aligned MHD equations 
  using 
  two different methods,
  a Lagrangian remap (Lare) approach, 
  as described for 3D MHD in
  \cite{paper:Arber2001}, adapted for 1D field-aligned
  hydrodynamics (Lare1D)
  and the adaptive mesh code HYDRAD
  \citep{paper:Bradshaw&Mason2003}.
  Time-splitting methods 
  are used in Lare to update thermal conduction and optically 
  thin radiation separately from the advection terms, 
  as discussed in Appendix \ref{app:A}. Furthermore, to treat 
  thermal conduction we use
  super time stepping (STS) methods, as         
  described in 
  \cite{paper:Meyeretal2012,paper:Meyeretal2014}  and 
  discussed in
  Appendix \ref{app:B}.
  \begin{figure}
    \centering
    \resizebox{\hsize}{!}
    {\includegraphics
    {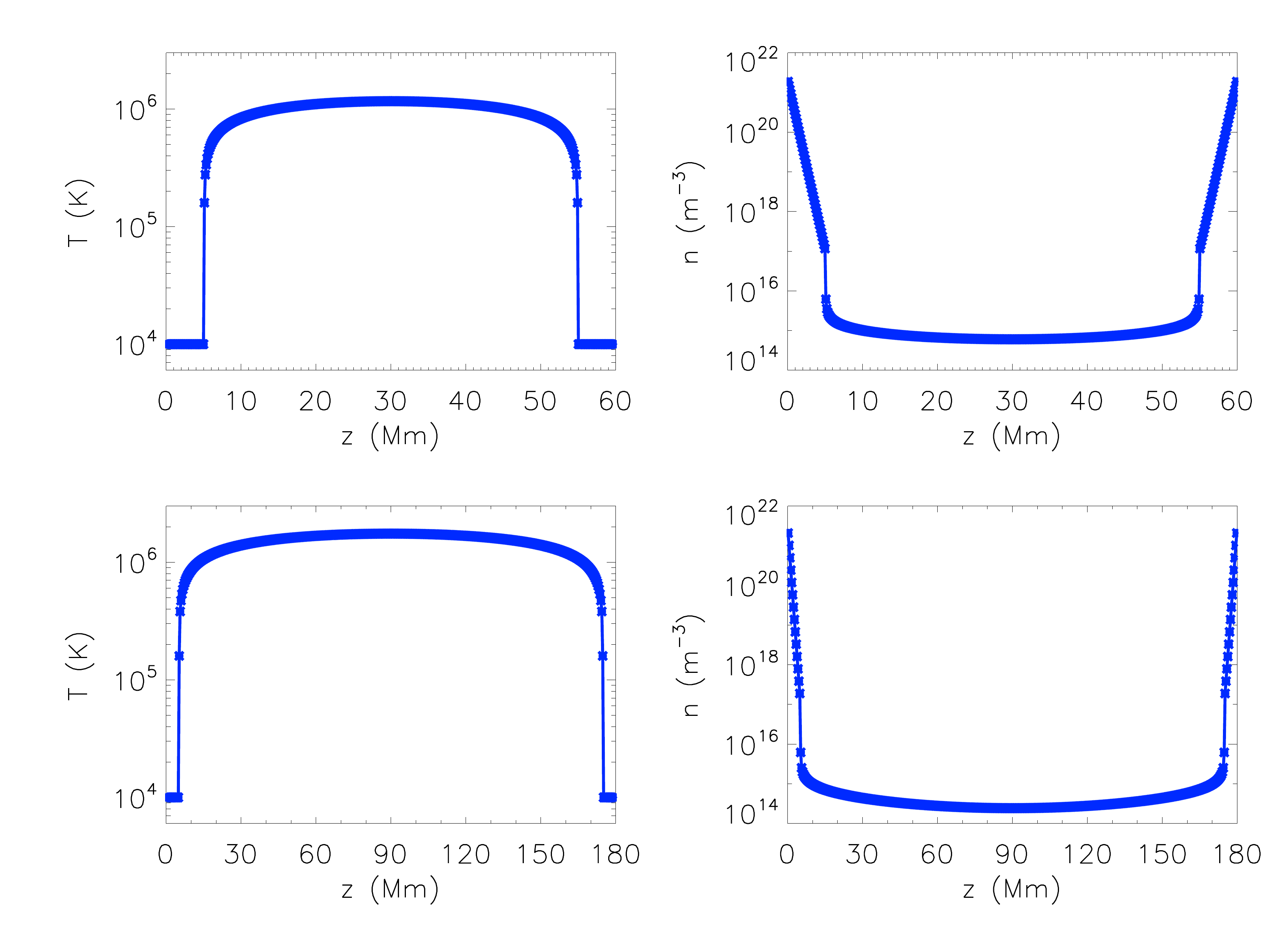}}
    \caption{Temperature and density initial conditions with 
    500 grid points along the length of the loop. Upper
    panels: $60$Mm loop 
    with $Q_{bg}=2.2167\times 10^{-5}$Jm$^{-3}$s$^{-1}$. 
    Lower panels:
    $180$Mm loop 
    with
    $Q_{bg}=6.8682\times 10^{-6}$Jm$^{-3}$s$^{-1}$.
    Each asterisk represents a single grid point.}
    \label{Fig:eq_nx=500_60Mm_180Mm_loop}
  \end{figure}
  \\
  \indent
  The initial condition of the model is a magnetic strand 
  (loop) in 
  static equilibrium. This is obtained by starting with an
  extremely high resolution uniform grid with $5\times10^5$ 
  grid points along the length of the loop. 
  We consider
  both a short (60Mm) and long (180Mm) loop, where the total    
  length 
  of each loop (2L) includes a 5Mm model
  chromosphere (included as a mass
  reservoir) at the base of 
  each TR ($z=5$Mm). 
  We set $T=10,000$K and n=$10^{17}$m$^{-3}$
  at the base of the TR.
  The initial equilibrium
  temperature and density profiles are then derived using 
  the same approach as described in
  \cite{paper:Bradshaw&Mason2003}.
  We note that, to achieve
  thermal balance, a
  small background heating term is necessary ($Q_{bg}$). 
  These fully resolved equilibrium solutions are then 
  interpolated onto the much coarser grids used for the 
  time-dependent evolution. 
  The initial conditions, with 500 grid points along the 
  length 
  of the loop, are shown for both the short and 
  long 
  loop in Fig.
  \ref{Fig:eq_nx=500_60Mm_180Mm_loop}.
  We note that neither solution is numerically resolved below 
  approximately $2\times 10^5$K until the chromospheric
  temperature is reached.
  
%
%

\subsection{Definitions}
\begin{figure}
    \centering
    \resizebox{\hsize}{!}
    {\includegraphics{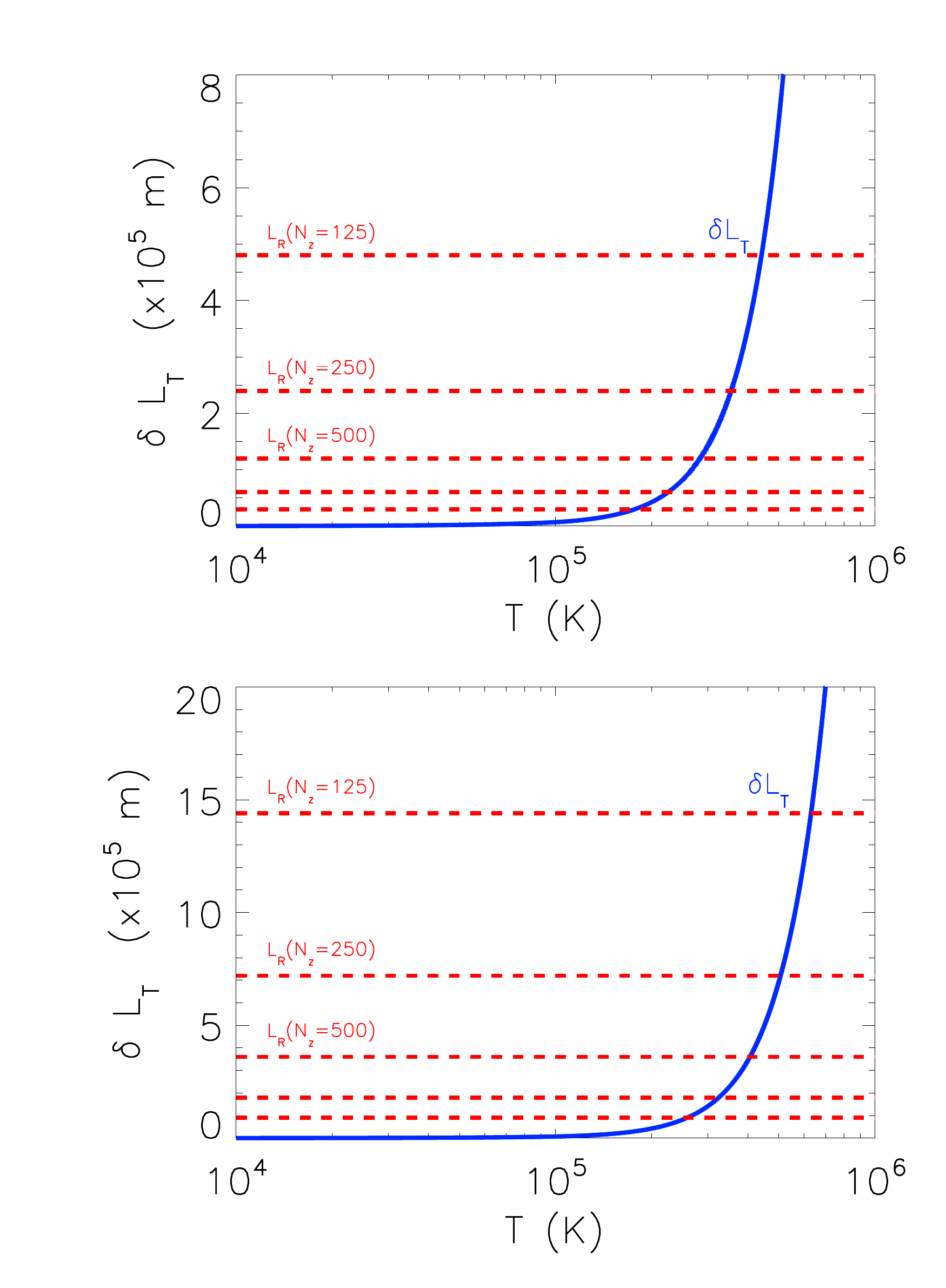}}
    \caption{
    \label{Fig:delta_L_T}    
    The product of $\delta$ and the 
    temperature length scale ($L_T$) as a function of 
    temperature (solid blue 
    line) 
    based on fully resolved equilibrium solutions 
    that are
    computed with $5\times 10^5$ grid points along the length 
    of the loop
    and are
    consistent with the short and long loop initial conditions 
    shown in Fig. \ref{Fig:eq_nx=500_60Mm_180Mm_loop}. 
    Upper panel: 
    $60$Mm
    loop.
    Lower panel: 
    $180$Mm loop. The dashed 
    red lines are the simulation resolutions ($L_R$) obtained 
    by using  
    different numbers of grid points. 
    In both plots, 
    starting from the top 
    the first dashed red line corresponds to 125 grid points, 
    the second to 250 grid points, the third to 500 grid
    points, the fourth to 1,000 grid points and the fifth to
    2,000 grid points. 
      }
  \end{figure}

  \indent
  We use coarse spatial resolutions and address the influence 
  of poor numerical resolution
  by modelling the unresolved region of the atmosphere,
  which we refer to as the UTR,
  as a discontinuity by using an appropriate jump condition,
  instead of trying to implement a grid that
  fully resolves the TR.
  To facilitate the formulation of this approach, we first 
  introduce
  some definitions.
  We define the temperature length scale as,
  \begin{align}
    \hspace{2cm}
    L_T = \frac{T}{|dT/dz|} = \frac{\kappa_0 T^{7/2}}{|F_c|}.
    \label{eqn:1d_L_T} 
  \end{align}
  With a uniform grid,
  the resolution
  in the simulation is given by,
  \begin{align}
    \hspace{3cm}
    L_R = \frac{2L}{N_z-1},
    \label{Fig:1d_L_R}
  \end{align}
  where $N_z$ is the number of grid points along the length of 
  the loop (2L). (A non-uniform grid will have the same 
  problems, 
  amenable with a similar solution.)
  Using these definitions, 
  we define the top of the UTR 
  ($z_0$) to be the final location, when 
  moving downwards from the loop apex ($z_a$), at which the 
  criteria,
  \begin{align}
    \hspace{3cm}
    \frac{L_R}{L_T} \leq \delta < 1,
    \label{eqn:1d_T_res_criteria} 
  \end{align}
  is satisfied. To ensure that we 
  have 
  sufficient resolution at the top of this region, that is   
  multiple grid points across the 
  temperature length scale, we take $\delta=1/4$
  throughout this paper. 
  \\
  \indent
  Fig. \ref{Fig:delta_L_T} demonstrates the consequences of 
  Eq. \eqref{eqn:1d_T_res_criteria} for short (long) 
  loops in 
  the upper (lower) panel. The product of
  $\delta $ and $L_T$ is shown as a function of temperature 
  (solid blue line) 
  with 
  the red dashed lines showing different values of $L_R$. Any 
  temperature that falls below the dashed lines will be part 
  of an UTR. This arises below 
  a few $10^5$K. 
  Also when coarse resolution is used, the 
  temperature at the
  top of the UTR is only weakly 
  dependent on the spatial resolution.
  \\
  \indent
  Lastly, 
  we define the base of the TR ($z_b$)
  to be the location at which the temperature first reaches or 
  falls below the chromospheric temperature (10,000K). 
  Employing these definitions it is straightforward
  to locate both the top of the UTR
  and the base of the TR at all 
  time steps during a simulation.   
  %
  %

%
%

\section{Unresolved transition region jump condition
  \label{section:Unresolved Transition Region Jump Condition}}
  \indent
  On use of equations
  \eqref{eqn:1d_continuity}--\eqref{eqn:1d_tee}, 
  one can write an equation for the 
  total energy in conservative form,
  \begin{align}
    \frac{\partial E }{\partial t} 
    =-
    \frac{\partial}
    {\partial z} (Ev + Pv + F_c) + Q - n^2 \Lambda(T),
    \label{eqn:1d_tec}
  \end{align}
  where the total energy is the sum of thermal, kinetic and 
  gravitational potential energy,
  \begin{align}
    \hspace{2cm}
    E = \frac{P}{\gamma - 1} + \frac{1}{2} \rho v^2 
    + \rho \Phi.
    \label{eqn:1d_te}
  \end{align}
  Here, $\Phi$ is the gravitational potential
  ($g_\parallel=d\Phi/dz$).
  \\
  \indent
   We integrate Eq.
  \eqref{eqn:1d_tec} over the UTR
  (of length $\ell$),
  from the base of the TR ($z_b$) upwards to the 
  top of the UTR ($z_0$), to obtain, 
  \begin{align}
    \ell \frac{d\bar{E} }{dt} 
    =&-E_0v_0
    -P_0v_0
    -F_{c,0}
   \nonumber
    \\
    &+E_bv_b
    +P_bv_b
    +F_{c,b}    
    + \ell \bar{Q} - \mathcal{R}_{utr},
    \label{eqn:1d_si_utr_tec}
  \end{align}
  where the 
  subscripts 0 and b indicate  quantities evaluated at the
  top and base of the UTR, respectively. The overbars 
  indicate spatial averages over the UTR and
  $\mathcal{R}_{utr}$
  is the integrated radiative losses (IRL)
  in the UTR (see Eq. \eqref{eqn:1d_R_utr} for
  the definition). 
  \\
  \indent
  Using the fully resolved HYDRAD results, we have confirmed
  that $F_{c,b}$ is 
  always small 
  ($F_{c,b}<<F_{c,0}$)
  and that after the intial downward 
  motion of 
  the TR (during the heating phase), the terms 
  containing $v_b$ are also significantly smaller
  than the remaining terms on the
  right-hand side (RHS) of  Eq.
  \eqref{eqn:1d_si_utr_tec}.
  It is these remaining terms that control the coronal
  response.
  Hence,
  we follow \cite{paper:Cargilletal2012a} and neglect these 
  terms from now on. 
  \\
  \indent
  We have also confirmed,
  from the fully resolved results, that 
  there are only short intervals (at the start of the heating 
  period) when $\ell d\bar{E}/dt$ can be significant. However, 
  the problem with including this term is that, with the 
  resolution of
  current 3D MHD models, it is very difficult to calculate 
  $\ell d\bar{E}/dt$
  accurately because the calculation requires $dE/dt$ to be
  integrated
  across the UTR.
  If the TR is not fully resolved then the heat flux 
  jumps across the UTR, resulting in the 
  estimates of $dE/dt$ being in error.
  Indeed, if we could
  calculate
  $\ell d\bar{E}/dt$ accurately,
  with coarse spatial resolutions, then it would not be 
  necessary to implement a method to
  obtain the correct upflow and evaporation.
  Therefore, the final assumption in the
  derivation of our jump condition is to adopt the 
  approach of \cite{paper:Klimchuketal2008} and
  neglect the left-hand 
  side (LHS) of Eq. 
  \eqref{eqn:1d_si_utr_tec}.
  \\
  \indent
  Under these assumptions, by combining equations 
  \eqref{eqn:1d_te} \& 
  \eqref{eqn:1d_si_utr_tec}, 
  we obtain the jump 
  condition at the top of the UTR, 
  \begin{align}
    \frac{\gamma}{\gamma - 1} P_0 v_0 + \frac{1}{2} \rho_0 
    v_0^3
    + \rho_0 \Phi_0 v_0
    =
    -F_{c,0}
    + \ell \bar{Q}
    - \mathcal{R}_{utr},
    \label{eqn:1d_utr_jc}
  \end{align}
  where
  \begin{figure}
    \centering
    \resizebox{0.7\hsize}{!}
    {\includegraphics{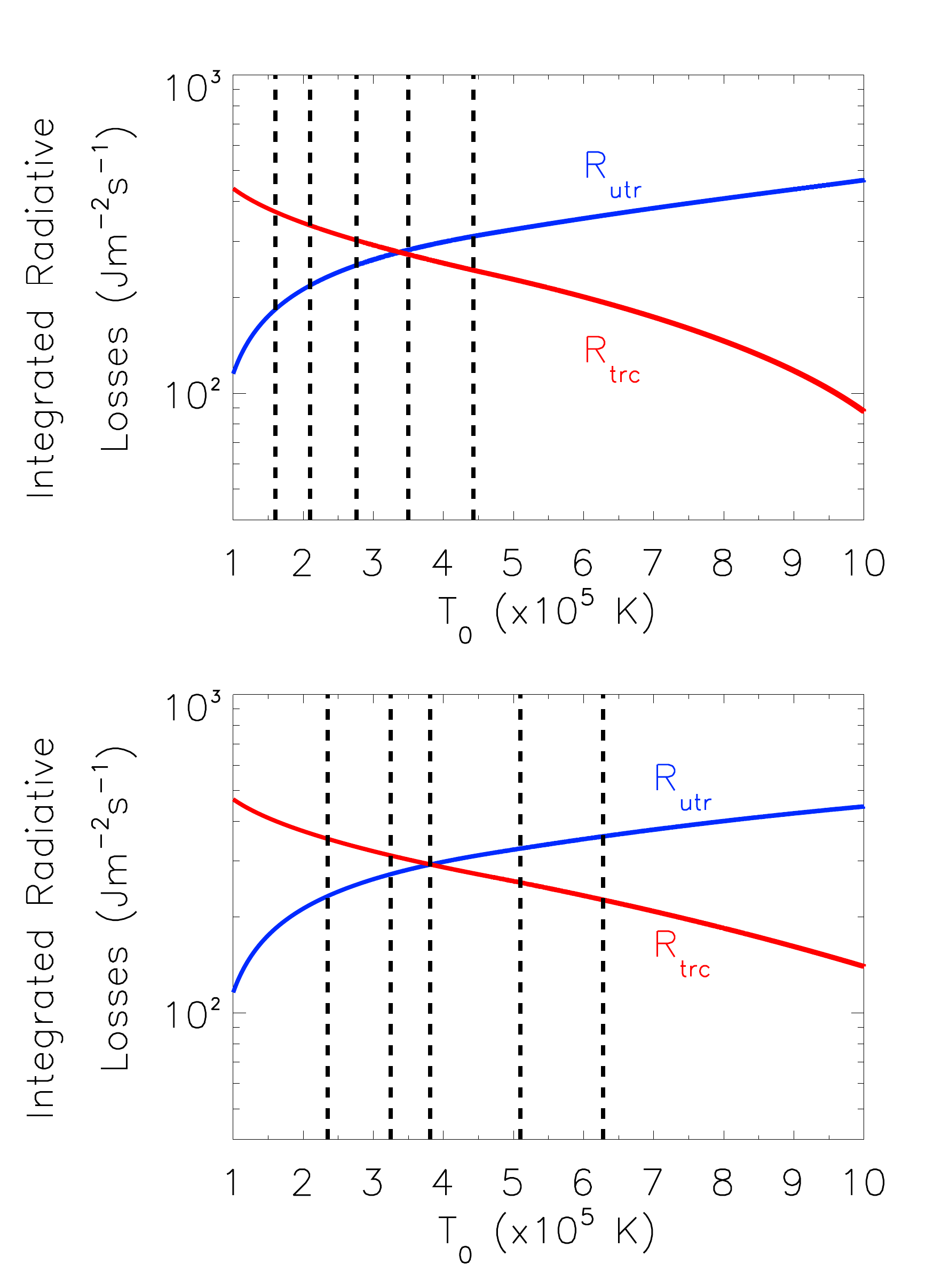}}
    \caption{    
    \label{Fig:integrated_radiative_losses_eq}
    IRL in the UTR (solid blue line) and resolved 
    upper 
    TR and corona (solid red line) 
    based on fully resolved equilibrium solutions 
    that are
    computed with $5\times 10^5$ grid points along the length 
    of the loop
    and are
    consistent with the short and long loop initial conditions 
    shown in Fig. \ref{Fig:eq_nx=500_60Mm_180Mm_loop}.  
    Upper panel: 
    $60$Mm
    loop.
    Lower panel: 
    $180$Mm loop.
    The dashed 
    black lines are the temperatures at the
    top of the UTR
    ($T_0$) that are obtained by using the  
    different simulation resolutions ($L_R$). 
    In both plots,
    starting from the right  
    the first dashed black line corresponds to 125 grid 
    points, 
    the second to 250 grid points, the third to 500 grid
    points, the fourth to 1,000 grid points and the fifth to
    2,000 grid points.  
    }
  \end{figure}
  the terms on the LHS are the 
  enthalpy flux ($ F_{e}$), kinetic energy flux and 
  gravitational 
  potential energy flux, respectively. The terms on the 
  RHS are the heat flux, 
  the average volumetric heating rate per unit cross-sectional 
  area
  and
  the IRL in the UTR respectively.
  We refer to Eq. \eqref{eqn:1d_utr_jc} as the 
  UTR jump condition and
  propose that the 
  UTR should be modelled as a discontinuity 
  using Eq. \eqref{eqn:1d_utr_jc} to impose a corrected 
  velocity ($v_0$) 
  at the 
  top of the UTR, at each
  time step.
  \\
  \indent
  This 
  corrected
  velocity is imposed following the conduction and 
  radiation and heating steps, prior to the advection step,
  as illustrated in Fig. \ref{Fig:Lare_update}, 
  while the flow at the base of the TR $
  (v_b)$ is subsequently
  accounted for during the advection step.
  Consequently, 
  at the time  
  of  calculation of $v_0$, it is possible to 
  calculate
  the 
  heat flux ($F_{c,0}$) and the average volumetric heating 
  rate 
  per unit   
  cross-sectional area in the UTR 
  ($\ell \bar{Q}$).
  Of the terms on the LHS of the UTR jump condition
  \eqref{eqn:1d_utr_jc}, 
  the pressure ($P_0$), density ($\rho_0$), and 
  gravitational potential ($\Phi_0$) are also known.
  The main challenge is the calculation of the 
  IRL in the UTR
  ($\mathcal{R}_{utr}$).
  \begin{table*}
    \caption{
    \label{table:simulations}
    A summary of the parameter space used and results from
    the numerical 
    simulations.} 
    \centering
    \resizebox{\hsize}{!}
    {
    \begin{tabular}{lccccccccc}
    \hline\hline
    \\
    Case & 2L & $Q_H$ & $\tau_H$ &
    $T_{\textrm{max}}(\textrm{HYDRAD})$ &
    $T_{\textrm{max}}(\textrm{LareJ})$ &
    $T_{\textrm{max}}(\textrm{Lare1D(500)})$ &
    $n_{\textrm{max}}(\textrm{HYDRAD})$ &
    $n_{\textrm{max}}(\textrm{LareJ})$ &
    $n_{\textrm{max}}(\textrm{Lare1D(500)})$ 
    \\
    & (Mm) & (Jm$^{-3}$s$^{-1}$) & (s) & (MK) & 
    (MK) & (MK) & ($10^{15}$m$^{-3}$) & ($10^{15}$m$^{-3}$) &
    ($10^{15}$m$^{-3}$)
    \\
    \hline
    1  & 60  & $8\times10^{-4}$ & 60  & 1.9 & 2.1
       & 2.1 & 0.86             &0.92 &0.74
    \\
    2  & 60  & $8\times10^{-3}$ & 60  & 5.7 & 6.1  
       & 6.1 & 2.2              & 2.6 & 1.5
    \\
    3  & 60  & $8\times10^{-2}$ & 60  & 12.5& 12.9
       &13.1 & 9.0              &11.6 &4.9
    \\
    4  & 60  & $8\times10^{-4}$ & 600 & 3.4 & 3.5
       & 3.5 & 2.2              & 2.6 & 1.0
    \\
    5  & 60  & $8\times10^{-3}$ & 600 & 6.9 & 7.1
       & 6.9 & 9.1              &11.4 & 2.9
    \\
    6  & 60  & $8\times10^{-2}$ & 600 & 13.7& 14.1
       &13.8 & 40.3             &49.7 &10.4
    \\
    7  & 180 & $5\times10^{-5}$ & 60  & 1.8 & 1.8
       & 1.8 & 0.28             &0.29 &0.27
    \\
    8  & 180 & $5\times10^{-4}$ & 60  & 2.9 & 3.1
       & 3.1  & 0.37            &0.40 & 0.33
    \\
    9  & 180 & $5\times10^{-3}$ & 60  & 9.3 & 10.2
       &10.2 & 1.0              &1.13 &0.42
    \\
    10 & 180 & $5\times10^{-5}$ & 600 & 2.5 & 2.7
       & 2.7 & 0.36             &0.40 &0.33
    \\
    11 & 180 & $5\times10^{-4}$ & 600 & 5.7 & 6.0
       & 6.0 & 0.98             &1.18 &0.39
    \\
    12 & 180 & $5\times10^{-3}$ & 600 & 12.3& 12.7
       &12.3 & 4.2              & 5.4 & 1.2
    \\
    \hline
    \end{tabular}
    }
    \tablefoot{
    The columns show
    the total length of the loop, the peak heating rate,
    the duration of the heating pulse and the maximum averaged
    temperature and density attained by
    HYDRAD
    (in single fluid mode) with the
    largest grid cell of width $400$km
    and 12 levels of refinement employed,
    and
    Lare1D
    using 500 grid points along the length 
    of the loop (coarse resolution)
    employed with (LareJ) and without (Lare1D(500)) the jump 
    condition, respectively.
    }
  \end{table*}
  %
  %
  
%
%

\subsection{Integrated radiative losses in the unresolved
  transition region}
  \indent  
  Motivated by equilibrium results, 
  we estimate 
  $\mathcal{R}_{utr}$ using the
  IRL in
  the resolved upper TR and corona 
  ($\mathcal{R}_{trc}$),
  \begin{align}
    \hspace{2cm}
    \mathcal{R}_{utr} = \int_{z_b}^{z_0}
    \! \! \! \!
    n^2 \Lambda(T)  \, dz \approx 
    \mathcal{R}_{trc},
    \label{eqn:1d_R_utr}
  \end{align}
  where,
  \begin{align}
    \hspace{2cm}
    \mathcal{R}_{trc} = 
    \int^{z_{{a}}}_{z_0} \! \! \! \! n^2 \Lambda(T) 
     \, 
    dz .
    \label{eqn:1d_R_trc}
  \end{align}
  \\
  To demonstrate the justification of 
  \eqref{eqn:1d_R_utr},
  in Fig. \ref{Fig:integrated_radiative_losses_eq} we plot  
  the IRL in
  the UTR 
  ($\mathcal{R}_{utr}$) and 
  resolved upper TR and corona
  ($\mathcal{R}_{trc}$) as functions of 
  the temperature at the top of the UTR ($T_0$), for both our 
  short and long loop
  initial conditions.
  These curves are obtained by integrating 
  the radiative losses from fully resolved solutions
  (using $5\times 10^5$ uniformly spaced grid points) 
  while adjusting the integration limits so that the
  spatial location of the 
  top of the UTR changes with the 
  temperature at this location.
  Previously, we have seen that when    
  coarse resolution is used, the 
  temperature at the
  top of the UTR is only weakly 
  dependent on the spatial resolution
  (see Fig. \ref{Fig:delta_L_T}), which means that there is 
  only a 
  small 
  range of resolvable TR temperatures before the unresolved 
  region of the atmosphere is reached,
  and within these small
  temperature ranges there is reasonably good  agreement 
  between the 
  values of 
  $\mathcal{R}_{trc}$ and $\mathcal{R}_{utr}$. For example,
  as can be seen
  in Fig. \ref{Fig:integrated_radiative_losses_eq},
  when using 1,000 grid points with $2L=180$Mm, 
  $T_0=3.25\times 10^5$K and
  $\mathcal{R}_{utr} \, (272$Jm$^{-2}$s$^{-1})$ 
  $\approx$
  $\mathcal{R}_{trc} \, (312$Jm$^{-2}$s$^{-1})$. We note that
  the agreement is even better when using 500 grid points.
  \\
  \indent
  But when coarse 
  resolution is used, a single grid point
  lower down in
  the atmosphere can have a
  considerable effect on the IRL in 
  the
  resolved upper TR and corona. Therefore, we note that it 
  is safer to define the top of the UTR to be a few grid cells 
  higher up than previously defined.

%
%

\subsection{Implementation of the jump condition}
  \indent  
  Once the IRL in the UTR ($\mathcal{R}_{utr}$) have been 
  estimated, 
  the
  corrected velocity ($v_0$) is then calculated, by
  firstly 
  solving the UTR jump condition \eqref{eqn:1d_utr_jc},
  which is a cubic in
  $v_0$,
  using a simple Newton-Raphson solver with the
  starting
  condition,
  \begin{align}
    \hspace{2cm}
    v_i
    =
    \dfrac{
    -F_{c,0}
    + \ell \bar{Q}
    - \mathcal{R}_{utr}}
    {(\gamma - 1)/\gamma P_0},
  \end{align}
  which is obtained by neglecting the kinetic energy 
  and gravitational potential energy fluxes in Eq. 
  \eqref{eqn:1d_utr_jc}.
  Convergence to a solution of the complete equation
  is rapid. 
  \\
  \indent
  In some cases approximation \eqref{eqn:1d_R_utr}
  underestimates the IRL in the 
  UTR, which may lead to spurious supersonic upflows for the 
  class of problems considered in this paper. Therefore, the 
  solution to Eq.
  \eqref{eqn:1d_utr_jc}, 
  $\tilde{v}_{0}$,
  is adjusted by
  using the following sound speed limiter,
  \begin{align}
    \hspace{3cm}
    v_0= \dfrac{\quad \tilde{v}_{0} \times c_s }
    {\sqrt{\tilde{v}_{0}^2 + c_s^2}},
  \end{align}
  where $c_s$ is the local sound speed at the top of the UTR. 
  It is this adjusted velocity ($v_0$) that we 
  impose
  at the
  top of the
  UTR. 
  This is consistent with the corresponding fully resolved 
  loop simulations (that 
  use an adaptive
  mesh), since no supersonic flows are present
  at the location where the jump condition is
  implemented, in all of the 12 cases considered. Hence, this 
  approximation is
  satisfactory for the problems presented here and it does not 
  inhibit 
  the existence of 
  supersonic flows higher up in the atmosphere.

%
%
  %
  %
  \begin{table}
    \caption{
    \label{table:ct}
    Numerical simulation computation times.}
    \vspace{-0.25cm}
    \centering
    \resizebox{\hsize}{!}
    {
    \begin{tabular}{lccccc}
    \hline\hline
    Case & $\tau(\textrm{\scriptsize LareJ})$
    & $\tau(\textrm{\scriptsize HYDRAD})$
    & $\tau(\textrm{\scriptsize Lare1D(8,000)})$ 
    & $\tau(\textrm{\scriptsize HYDRAD})/$
    & $\tau(\textrm{\scriptsize Lare1D(8,000)})/$ 
    \\
    & (mins) & (mins) & (mins) 
    & $\tau(\textrm{\scriptsize LareJ})$
    & $\tau(\textrm{\scriptsize LareJ})$
    \\
    \hline
    1  & 17  & 316   & 7,426  & 18.6 & 436.8
    \\
    2  & 19  & 340   & 7,766  & 17.9 & 408.7
    \\
    3  & 51  & 1,943 & 13,886 & 38.1 & 272.3
    \\
    4  & 22  & 370   & 6,341  & 16.8 & 288.2
    \\
    5  & 82  & 2,617 & 8,594*  & 31.9 & 106.0
    \\
    6  & 154 & 5,177 & 12,732* & 33.6 & 82.7
    \\
    7  & 26  & 1,559 & 18,893 & 60.0 & 726.7
    \\
    8  & 28  & 1,566 & 18,059 & 56.0 & 645.0
    \\
    9  & 35  & 1,605 & 16,833 & 45.9 & 480.9
    \\
    10 & 26  & 1,805 & 11,138 & 69.4 & 428.4
    \\
    11 & 32  & 1,914 & 11,997 & 59.8 & 374.9
    \\
    12 & 86  & 2,269 & 12,973* & 26.4 & 150.8
    \\
    \hline
    \end{tabular}
    }
    \vspace{-0.25cm}
    \tablefoot{
    The columns show
    the computation times 
    (run on a single processor) using the Lare1D 
    code
    with 500 grid points (coarse resolution)
    employed with the jump 
    condition (LareJ),    
    the HYDRAD code
    (in single fluid mode) with the
    largest grid cell of width $400$km
    and 12 levels of refinement employed,
    the Lare1D code using 8,000 grid points along the length 
    of the loop (Lare1D(8,000)), and the computational
    time ratios between these methods.
    The short loop simulations (Cases 1-6) are run to a final 
    time of 
    4,000s and the long 
    loop simulations (Cases 7-12) are run to a final 
    time of 
    12,000s. The asterisks indicate cases where the Lare1D
    code using 8,000 grid points was unable to resolve
    the density to within 75\% of the HYDRAD solution.
    }
  \end{table}
  \begin{figure*}
    \centering
    \includegraphics[width=17cm]
    {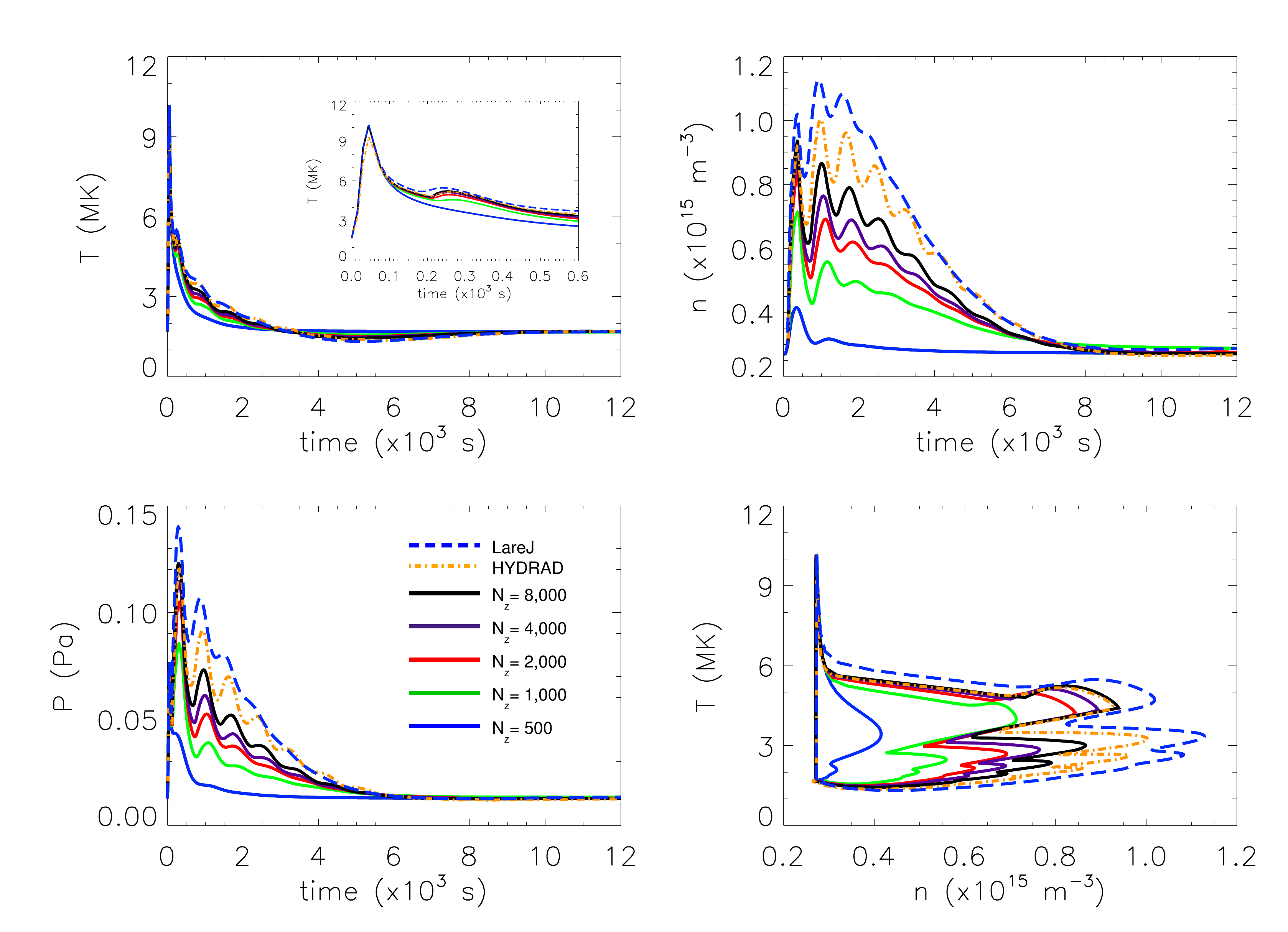}
    \caption{
    \label{Fig:g9_ca}
    Results for Case 9. The panels show the coronal averaged
    temperature, density and pressure as functions
    of time, and the temperature versus density phase space
    plot. 
    The solid lines represent the Lare1D solutions obtained
    by using different numbers of grid points along the
    length of the loop,  
    the dashed blue line is the LareJ solution (that
    is computed with the same spatial resolution as the
    solid blue curve) and the 
    dot-dashed orange line 
    corresponds to the HYDRAD solution.
    }
  \end{figure*}
  \begin{figure*}
    \centering
    \includegraphics[width=17cm] 
    {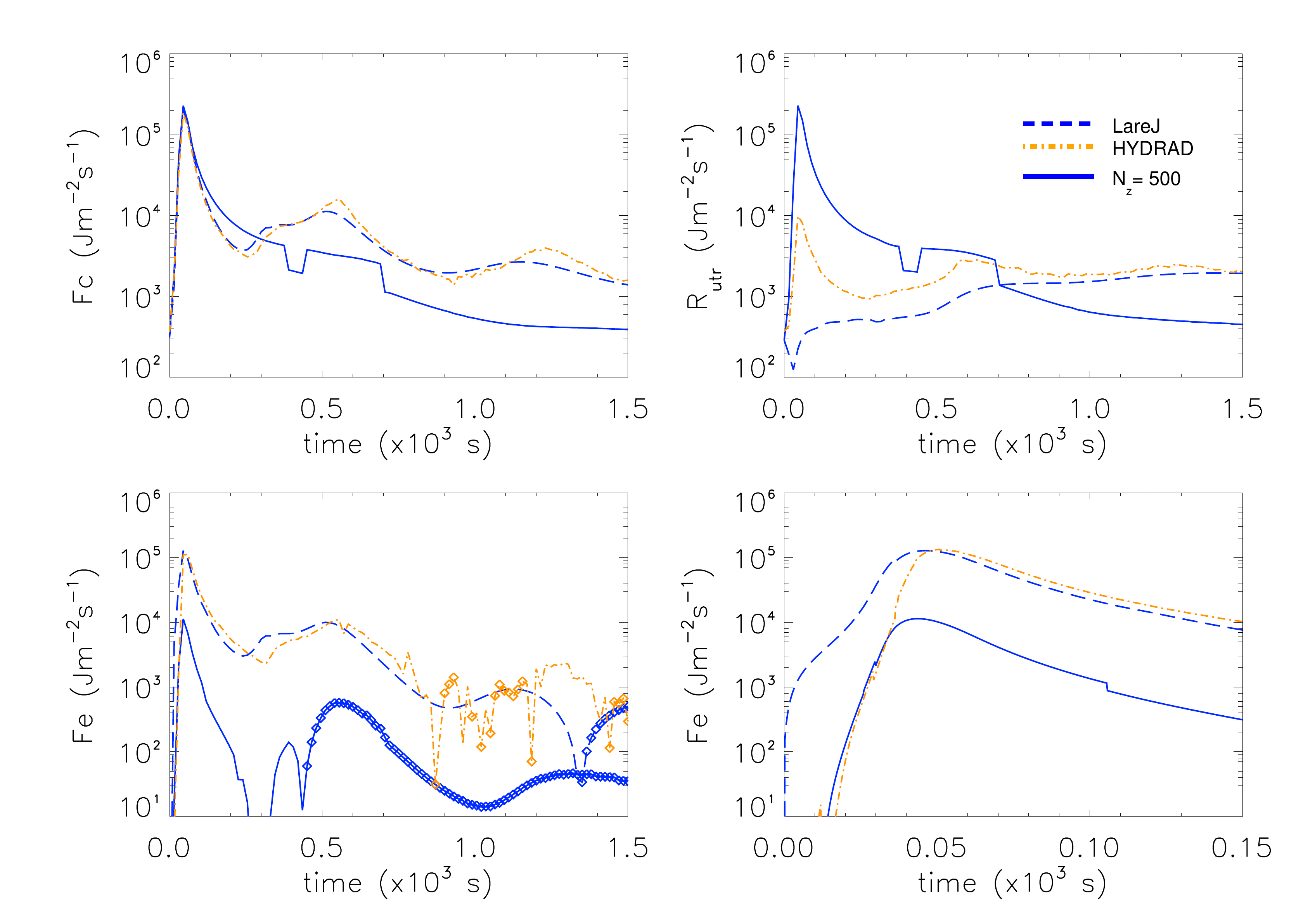}
    \caption{
    \label{Fig:g9_utr_terms}
    Results for Case 9. The panels show 
    the heat flux at the top of UTR,  
    IRL in the UTR
    and
    enthalpy flux over two different time intervals at the top 
    of the UTR
    (lines connected by diamond symbols indicate where 
    the enthalpy flux is downflowing and lines without 
    diamonds 
    indicate where the enthalpy flux is 
    upflowing) as functions of time.
    The dashed blue line is the LareJ solution
    (that
    is computed with 500 grid points along the length of 
    the loop), 
    the
    solid blue line
    is the Lare1D 
    solution that
    is computed with 500 grid points along the length 
    of 
    the loop and 
    the 
    dot-dashed orange line 
    corresponds to the HYDRAD solution.
    }
  \end{figure*}
  \begin{figure*}
    \centering
    \includegraphics[width=12cm]  
    {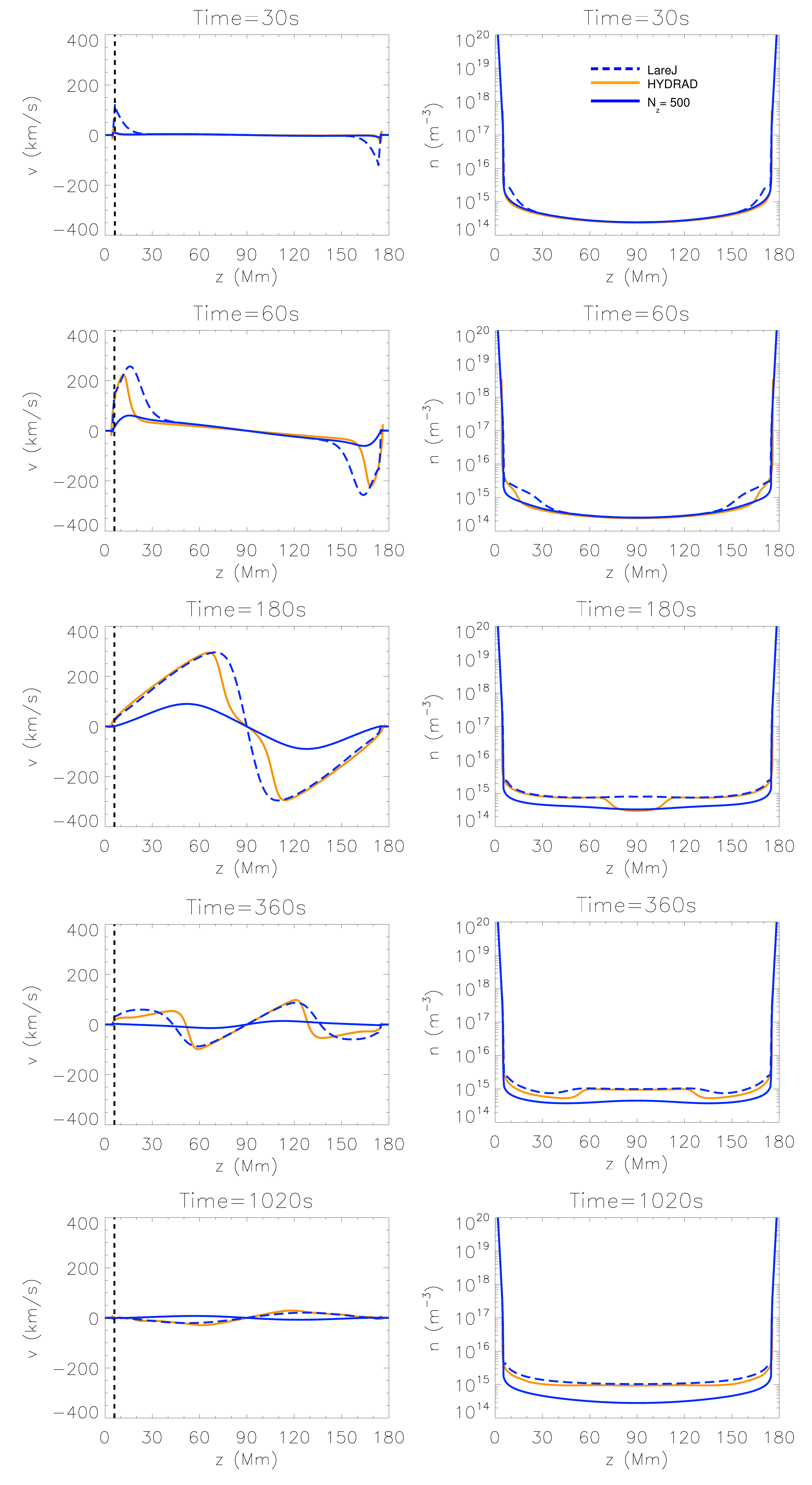}
    \caption{
    \label{Fig:g9_v_n_te}
    Results for Case 9. The panels show the velocity
    and density as functions of position for times during the 
    evaporation
    phase up until the second density peak.
    The dashed blue line is the LareJ solution
    (that
    is computed with 500 grid points along the length of 
    the loop), the
    solid blue line
    is the Lare1D 
    solution that
    is computed with 500 grid points along the length 
    of 
    the loop and 
    the 
    solid orange line 
    corresponds to the HYDRAD solution.
    The dashed black line indicates the position of the top of 
    the
    UTR ($z_0$)
    }
  \end{figure*}
  \begin{figure*}
    \centering
    \includegraphics[width=17cm]  
    {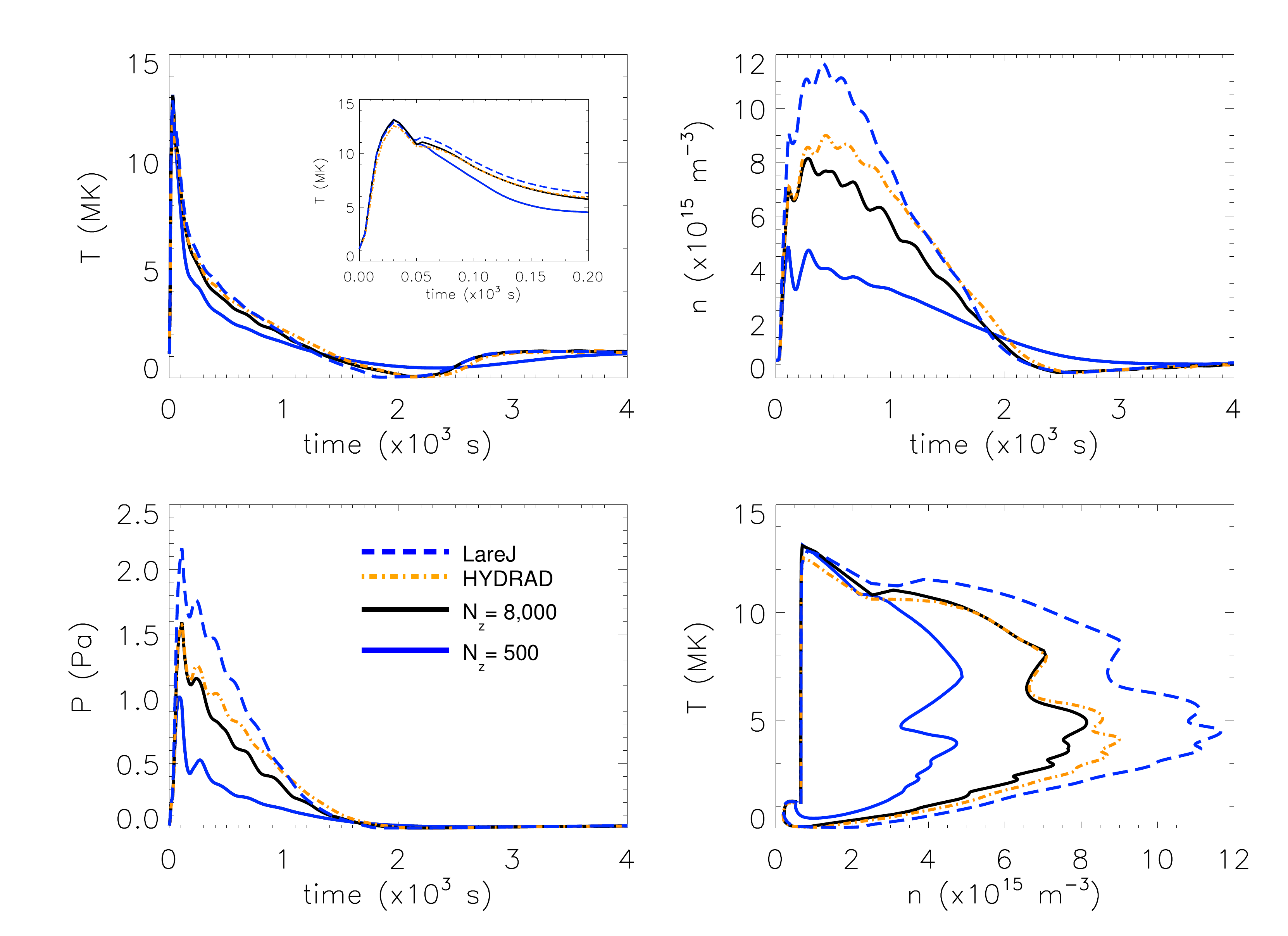}
    \caption{
    \label{Fig:g3_ca}
    Results for Case 3. 
    Notation is the same as Fig. \ref{Fig:g9_ca} but note
    the different time axis.
    }
  \end{figure*}
  \begin{figure*}
    \centering
    \includegraphics[width=17cm] 
    {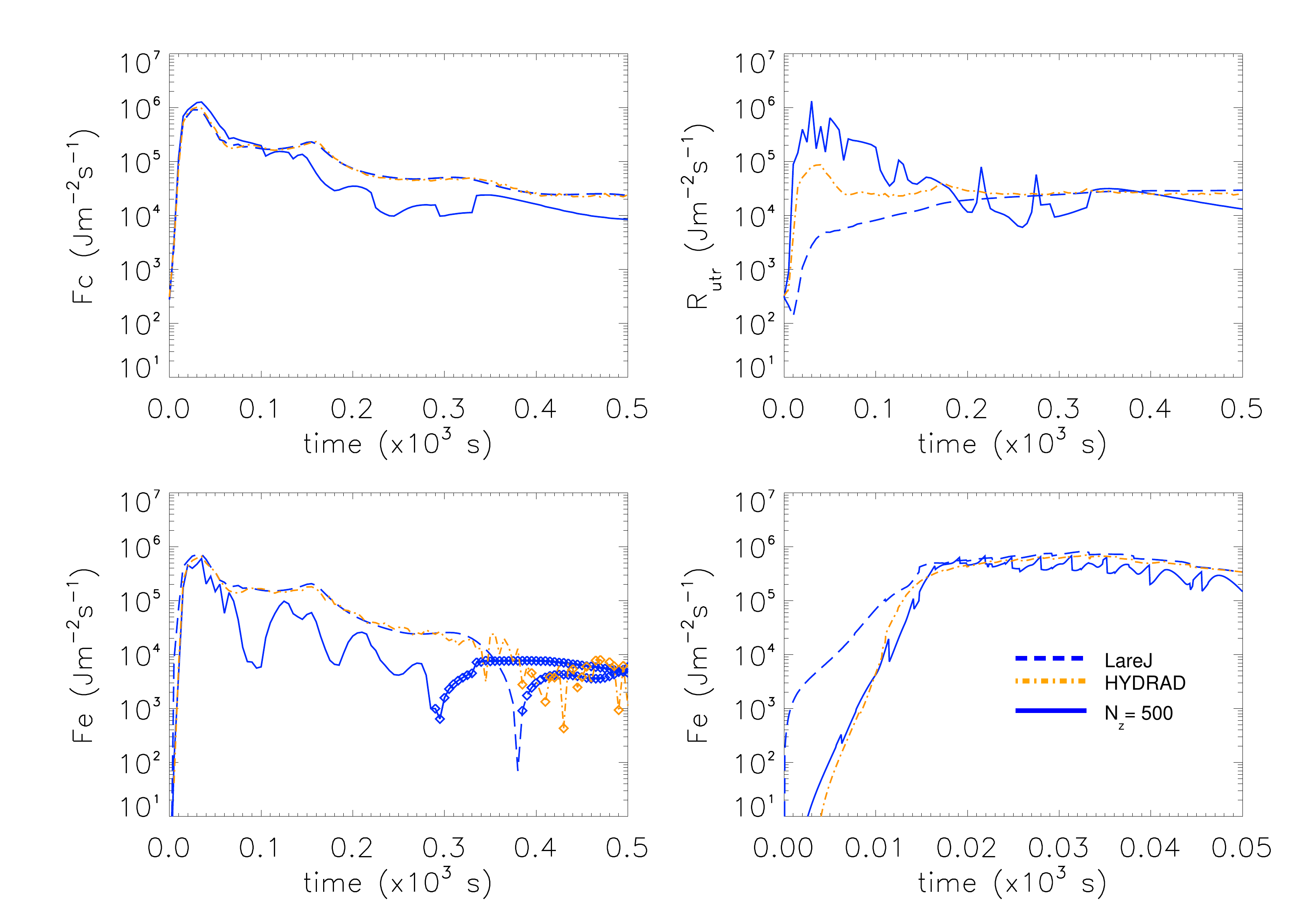}
    \caption{
    \label{Fig:g3_utr_terms}
    Results for Case 3. 
    Notation is the same as Fig. \ref{Fig:g9_utr_terms}
    but note
    the different time axis.
    }
  \end{figure*}
  \begin{figure*}
    \centering
    \includegraphics[width=16.5cm]  
    {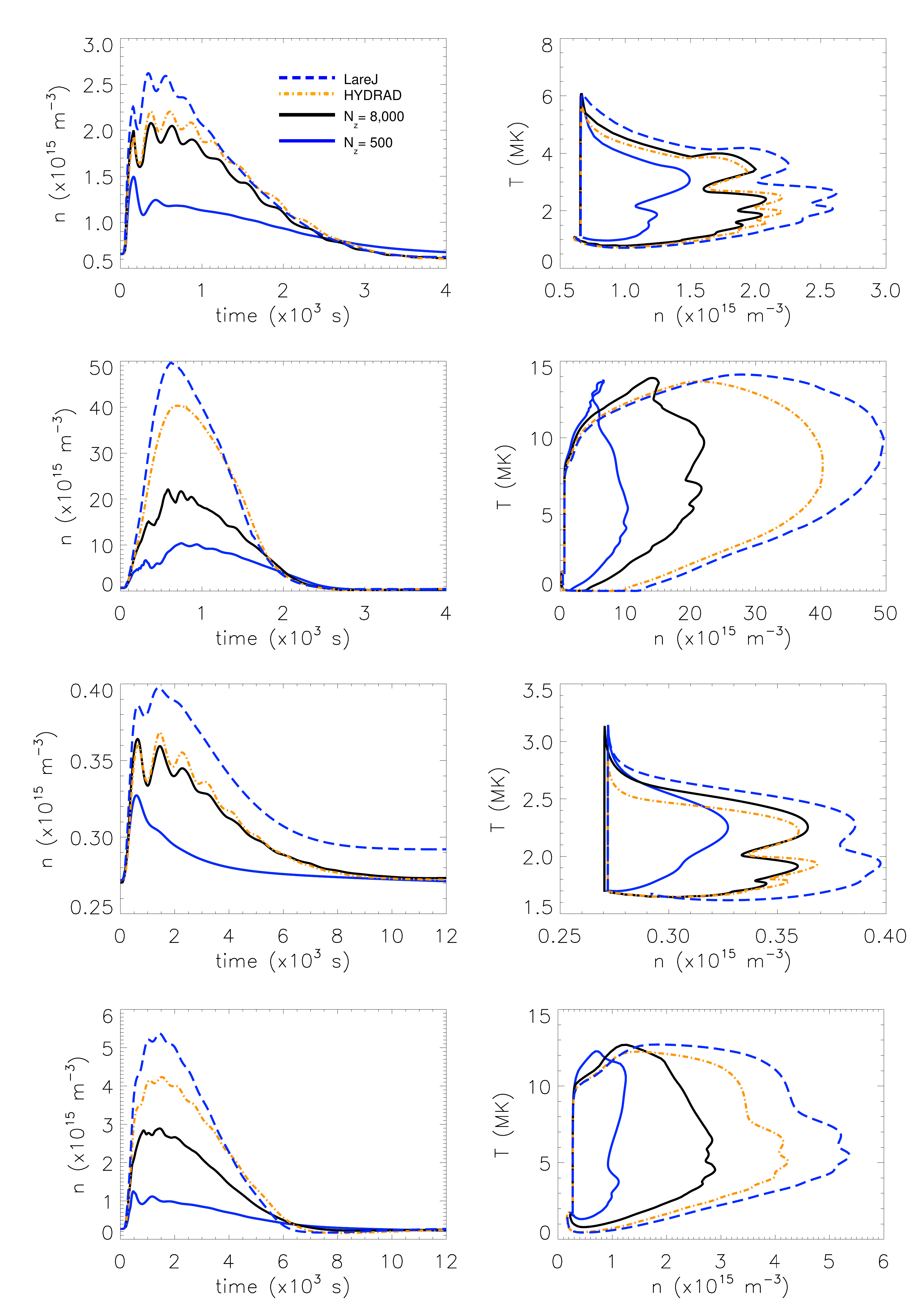}
    \caption{
    \label{Fig:g2_g6_g8_g12_ca}
    Results for Cases 2 (upper two panels), 6 
    (upper central two panels), 8 (lower central two panels) 
    and 12 (lower two panels).
    The panels show the coronal averaged density as
    a function
    of time and the temperature versus density phase space
    plot.
    }
  \end{figure*}
\section{Results
  \label{section:Results}}
  \indent
  The effectiveness of the UTR jump condition to
  obtain a physically realistic evolution, through the 
  complete 
  coronal heating and cooling cycle, when employed with 
  coarse resolution is 
  investigated for a series of impulsive coronal heating
  events. 
  The heating events considered 
  are based on the cases (1-12) 
  that were previously studied in
  BC13.
  These events
  are described in Table \ref{table:simulations} and cover 
  several orders of 
  magnitude and duration of heating for both a short and long 
  loop.
  The energy release is
  also the 
  same as that used in BC13. The temporal profile 
  is 
  triangular with a peak value of $Q_H$ and
  total duration of $\tau_H$  while
  the spatial profile is uniform along the loop.
  \\
  \indent 
  For each case, the main   
  assessment of the performance of the UTR jump condition 
  model is 
  a 
  comparison of Lare1D using 500 grid points
  employed with the jump 
  condition (referred to as LareJ), with both Lare1D without 
  the jump 
  condition but using up to 8,000 grid points and the 
  adaptive mesh code HYDRAD. The choice of 500 grid points is  
  motivated by 
  what 
  is routinely
  used in current multi-dimensional MHD models
  \citep{paper:Bourdinetal2013,
  paper:Hansteenetal2015,
  paper:Hoodetal2016,
  paper:Dahlburgetal2016}.
  The spatial resolution of these solutions is $120$km and 
  $360$km for the short and long loop, 
  respectively.
  \\
  \indent
  For the Lare1D 
  solutions we employ a 
  uniform grid and repeat each run with 
  $N_z=[500,\,1,000,\,2,000,\,4,000,\,8,000]$  grid points 
  along the length of the loop. We note that because we are
  using a uniform grid each time we double the number
  of grid points, even although we improve the TR resolution, 
  we also further reduce 
  the thermal conduction timescale in the corona and so the
  computational time increases. 
  Therefore, we have
  limited the most refined
  resolution used here because of the increased computation 
  time required. 
  \\
  \indent Consistent with our model equations 
  \eqref{eqn:1d_continuity}-\eqref{eqn:gas_law}, we run the
  HYDRAD code in single fluid mode. The HYDRAD code has an 
  adaptive grid that is capable of increasing the numerical 
  resolution wherever it is needed based on selected 
  refinement conditions. This enables the code to fully 
  resolve the small length scales in the 
  TR while retaining a coarser grid
  elsewhere.
  Following BC13,
  we select the largest grid cell to be of width $400$km and
  employ 12 levels of refinement, so that in the most highly 
  resolved regions the grid cells are of width $98$m. In this
  paper, we assume that the HYDRAD solution is 
  \lq correct\rq\ .
  
%
%
  \subsection{Case 9}
  \indent
  BC13 found their Case 9 
  (a strong nanoflare in a long loop) to be 
  one of   
  the more challenging examples for obtaining correct coronal 
  densities. Fig. \ref{Fig:g9_ca} shows
  the temporal evolution of 
  the coronal averaged temperature ($T$), density ($n$), 
  pressure ($P$) and the corresponding temperature versus
  density phase space plot.
  The coronal averages are computed by spatially averaging 
  over the uppermost 50\% of the loop. 
  (The trends are the same if either
  the averages are computed over 
  the full portion of the 
  loop above $z_0$ or the values are compared
  at the top of the UTR.)
  In the plots each solid line corresponds to a Lare1D 
  solution that was calculated by employing
  a different number of grid points along the length of the
  loop.
  The solid blue line has 500 grid points ($L_R\!=\!360$km), 
  the green line has 1,000 grid points ($L_R\!=\!180$km),
  the red line has 2,000 grid points ($L_R\!=\!90$km),
  the purple line has 4,000 grid points ($L_R\!=\!45$km)
  and the 
  black line has 8,000 grid points ($L_R\!=\!22.5$km).
  The dashed blue line is the LareJ solution that is 
  computed with
  500 grid points along the length of the loop and the 
  dot-dashed orange line 
  corresponds to the HYDRAD solution.
  \\
  \indent
  Starting with the Lare1D solutions it is clear that
  we recover the result presented by
  BC13, namely that
  the main effect of insufficient resolution is on the coronal 
  density while the temperature is far less resolution
  dependent. We also note that in this case, as is predicted 
  by 
  BC13
  the most refined 
  resolution that we employed with the Lare1D code is still 
  not 
  capable of reproducing the fully resolved HYDRAD solution.
  \\
  \indent
  However, if we focus on the LareJ 
  solution, 
  there is good agreement between the LareJ and HYDRAD
  solutions.
  At the initial 
  density peak, the LareJ solution
  evaporates about 10\%  too much material upwards into the 
  corona, in comparison to the HYDRAD solution, while 
  the density of the corresponding coarse
  Lare1D solution (run with the same spatial
  resolution, solid blue line) is more than a factor of two 
  lower 
  than the resolved loop value.
  As a consequence of this difference in densities, because
  the conductive cooling timescale scales 
  as $n/T^{5/2}$, 
  the LareJ solution cools at the correct rate while
  there is evidence that the corresponding coarse Lare1D 
  solution 
  cools more rapidly. 
  \\
  \indent
  The density then oscillates as the
  plasma sloshes to and fro within the loop. These
  oscillations are captured to a large extent by the LareJ 
  solution but are not prominent in the corresponding 
  coarse Lare1D solution. During 
  these oscillations, even although the LareJ density 
  remains slightly 
  too high, the accuracy of the
  LareJ solution
  is still an improvement 
  on even the most refined Lare1D solution. The LareJ 
  solution   
  then goes on to attain the correct draining rate during the
  density decay phase before recovering the
  equilibrium. 
  \\
  \indent
  Bringing all these factors together, in the phase space plot
  it is evident that the LareJ solution
  captures the evolution of the density as a function
  of temperature
  more accurately than the
  entire set of Lare1D solutions, including the most refined 
  solution that has a
  factor of 16 more grid points along the length of the loop.
  \\
  \indent
  Table \ref{table:ct} summarises the CPU requirements for all 
  cases and  
  demonstrates the large gain in CPU time of the UTR
  jump condition method 
  over the HYDRAD and most refined Lare1D runs. 
  Therefore, in this particular case,
  our method obtains a coronal 
  density comparable 
  to HYDRAD (fully-resolved 1D 
  model) but with a significantly faster computation time and 
  also provides a
  significant improvement in the accuracy of the 
  coronal density evolution when 
  compared to the equivalent simulations run without the
  jump condition. 
  \\
  \indent
  Using HYDRAD, 
  BC13 demonstrated that,
  for reasonably
  accurate solutions in the case of 180Mm loops and peak 
  temperatures exceeding 6MK,
  cell widths of no more than 5km are required. 
  What we have shown in Fig. 
  \ref{Fig:g9_ca} - \ref{Fig:g9_v_n_te} is that it is possible 
  to obtain realistic densities, temperatures and velocities
  with cell widths of 360km by using the UTR jump condition
  employed in LareJ.
  \\
  \indent
  We now turn our attention to understanding why the LareJ 
  solution performs well for this particular heating event 
  (Case 9). Fig. \ref{Fig:g9_utr_terms} shows the 
  temporal evolution of the heat and enthalpy
  fluxes at the top of the UTR and 
  the IRL in the
  UTR.
  These quantities are the
  dominant terms in the UTR jump condition
  \eqref{eqn:1d_utr_jc} 
  although the loop's 
  evolution can be influenced by the additional terms
  in Eq. \eqref{eqn:1d_utr_jc} 
  that are not 
  shown here.
  The dashed blue lines represent the appropriate LareJ 
  quantities 
  and the dot-dashed orange (solid blue) lines 
  represent the 
  appropriate quantities that are obtained throughout the
  evolution of the 
  HYDRAD 
  solution
  (Lare1D solution computed with 500 grid points
  along the length of the loop) .
  To calculate these quantities
  the definition of the UTR is determined based
  on the time evolution of the 
  temperature from the LareJ solution.
  \\
  \indent
  During the initial evaporation phase (first 400s) the 
  excess heat flux drives an upward enthalpy flux. 
  Throughout this phase there is 
  good agreement  between the enthalpy fluxes of
  the LareJ and 
  HYDRAD solutions. 
  This agreement is achieved because the
  downward heat flux dominates
  the IRL in the UTR and so the 
  UTR jump condition principally returns the 
  heat flux as an upward enthalpy flux.
  \\
  \indent
  However, close inspection reveals that,
  throughout the first 40s (see lower right panel in
  Fig. \ref{Fig:g9_utr_terms}), 
  the enthalpy flux of
  the LareJ solution exceeds that of the HYDRAD solution. 
  During this period the
  LareJ radiation approximation \eqref{eqn:1d_R_utr} is least 
  accurate
  and leads to an underestimation of the 
  IRL in the UTR. It is this 
  underestimation of 
  the IRL that drives the enhanced
  enthalpy flux.
  \\
  \indent
  Fig. \ref{Fig:g9_v_n_te} shows the
  velocity
  and density
  as functions of position,
  from the LareJ, coarse Lare1D and HYDRAD simulations,
  for times during the 
  evaporation
  phase up until the second density peak. The enhanced
  enthalpy flux, throughout the first 40s, indicates that the
  correcting velocity ($v_0$), 
  imposed
  at the top of the UTR, is overestimated during this period. 
  This is confirmed in the top left 
  panel in Fig.
  \ref{Fig:g9_v_n_te}. Therefore, the underestimation
  of the IRL in the UTR leads to an overestimation in the 
  initial upflow,
  locally at the top of the UTR, which then generates an 
  enhanced global velocity that facilitates the over
  evaporation of the LareJ solution.
  \\
  \indent
  Despite this overestimation in the 
  initial upflow, by imposing the 
  correcting velocity ($v_0$) locally at the top of UTR, 
  the jump condition method is still able to capture the  
  global 
  velocity much more accurately, in time, than the 
  corresponding
  simulation run without the jump condition
  (see Fig. \ref{Fig:g9_v_n_te}). 
  \\
  \indent 
  Radiation becomes increasingly important as
  the density increases. Then, at the time when the 
  radiation finally exceeds the heat flux,
  the loop enters the 
  density decay phase because a downward 
  enthalpy flux
  (condensation) is required to power the TR 
  radiation. During this decay phase,
  the LareJ solution drains material from
  the corona at the correct rate 
  due to the improvement in the accuracy 
  of 
  the LareJ radiation estimation
  \eqref{eqn:1d_R_utr}, following the first 
  density peak.
  
%
%
  \subsection{Case 3}
  BC13 found their Case 3
  (a small flare in a short loop)
  demanded the most severe requirements on the 
  spatial resolution.  
  Grid cells of width 390m were needed, in the most refined
  regions, in order for the coronal density to exceed 90\% of
  the properly resolved value. 
  The results for the numerical simulations included in this 
  case are shown in
  Fig. \ref{Fig:g3_ca} and \ref{Fig:g3_utr_terms}.
  To show the comparison exclusively
  between the key solutions,
  in the coronal averaged plots, we now drop the intermediate
  Lare1D solutions.
  \\
  \indent
  In this particular case, even although the LareJ solution
  suffers from its most significant over evaporation  
  at the initial density peak (about 30\%) and the 
  density remains too high throughout the first 1,000s, 
  its
  performance remains reasonably encouraging from
  the viewpoint that the LareJ solution 
  follows the same fundamental evolution as the HYDRAD 
  solution and their agreement is good throughout the 
  density decay phase.
  The factors responsible for driving this
  behaviour in the LareJ solution are the same as those 
  seen previously in Case 9.

%
%
  \subsection{Remaining cases}
  \indent
  We present
  the numerical comparison for the remaining cases
  in Table 
  \ref{table:simulations}, where
  the maximum averaged coronal 
  temperature and density attained by the
  HYDRAD, LareJ and corresponding coarse Lare1D 
  solutions are shown.
  In all 12 cases, the table shows that
  the accuracy of the maximum coronal density is 
  considerably improved with the LareJ solution
  when compared to the same resolution run 
  without the jump condition implemented.
  \\
  \indent
  The results for the Cases 2, 6, 8 and 12 are 
  shown in Fig. \ref{Fig:g2_g6_g8_g12_ca}.
  Essentially, because we drive the temperature 
  throughout 
  the impulsive heating event, we have seen that 
  the temporal evolution of the coronal averaged temperature
  is only weakly dependent on both the spatial resolution and
  computational method used. 
  Therefore, it is sufficient to now show only 
  the temporal evolution of 
  the coronal averaged density and the corresponding 
  temperature versus
  density phase space plots.
  \\
  \indent
  In these cases, the UTR jump condition method consistently 
  captures  
  a physically realistic evolution, through the 
  complete 
  coronal heating and cooling cycle, 
  comparable to that of the HYDRAD solutions. 
  The estimation of the 
  IRL in the UTR is again identified as the main source of 
  error that drives the observed over evaporation.
  This is due to the
  simple radiation estimation \eqref{eqn:1d_R_utr} used
  and despite this,
  it remains clear that as
  a first approximation, the LareJ solutions are reasonably 
  good, providing a significant improvement on the 
  corresponding coarse simulations run without the jump 
  condition.
  \\
  \indent
  However, we note from the phase space plot of Case 
  8 that (1) for this particular heating event,
  the most refined Lare1D solution (the 
  black line, computed with 8,000 grid 
  points) has a much better agreement with HYDRAD
  than the LareJ solution
  and (2) the 
  LareJ
  solution does not recover the exact long loop initial 
  equilibrium, but 
  returns to another nearby equilibrium
  with an increased density of
  around 7\% (similar behaviour was also seen in BC13).
  This is true for all of the long loop cases considered
  but is only observable in those where the density increase,
  in response to the heating event, is small
  (e.g. Cases 7 \& 8).

%
%

\section{Discussion and conclusions
  \label{section:Discussion and Conclusions}}
  \indent
 The difficulty of obtaining adequate spatial resolution in 
 numerical simulations of the corona, transition region (TR)
 and
 chromosphere system has been a long-standing problem. As 
 pointed out by BC13, the main consequence of not   
 resolving the TR is that the resulting coronal density is 
 artificially low. This paper has presented an approach to 
 deal with this problem by using an integrated form of energy 
 conservation that essentially treats the lower TR as a 
 discontinuity. Hence, the response of the TR to changing 
 coronal conditions is determined through the imposition of a 
 jump condition. When compared to fully resolved 1D 
 models 
 (e.g. BC13), our new 
 approach 
 generated improved coronal densities with significantly 
 faster computation times than the
 corresponding high-resolution and fully 
 resolved models. Specifically, our 
 approach required at least one to two orders of magnitude 
 less computational time than fully resolved (high-resolution) 
 models.
 \\
 \indent
 The 12 cases presented in this paper were selected to 
 correspond to the benchmark cases presented by BC13. 
 In 
 all 12 cases, the evolution of the coronal density is 
 considerably improved, compared to the same resolution run 
 without the jump condition implemented. Crucial here, is to 
 obtain a reasonable estimate of the (integrated) radiative 
 losses in the unresolved part of the TR. 
 \\
 \indent
 We have considered only spatially uniform impulsive heating   
 events.   
 Simulations with the heating concentrated
 either at the loop base or near the loop apex will be
 presented in a subsequent
 publication.
 \\
 \indent
 The advantages of this new approach are multiple. For 1D 
 hydrodynamic simulations of the coronal response to heating 
 \citep[see e.g.][for a review]{paper:Reale2014}, 
 the short computation 
 time means that (a) simulations of coronal heating events can 
 be run quickly, permitting an extensive survey of the (large) 
 parameter space and (b) simulations of multiple loop strands 
 (thousands or more) that either comprise a single observed 
 loop (e.g. a core loop), or an entire active region, can be 
 performed with relative ease. In 3D MHD codes, the method can 
 be included without the need for higher spatial resolution 
 and a corresponding extended computation time. Indeed, our 
 results suggest that good accuracy can be obtained with the 
 order of 500 grid points, typical of what is routinely used 
 in current 3D MHD simulations. The extension to 3D will be 
 addressed fully in a future publication.
  \\
  \indent
  The work presented here has adopted the simplest possible 
  model for the radiation in the lower, unresolved transition 
  region (UTR), 
  and leads to improved coronal densities.
  The estimate used
  was motivated by the 
  calculation of the radiation integrals for the equilibrium 
  conditions 
  (as shown in Fig. 
  \ref{Fig:integrated_radiative_losses_eq}), at which the 
  error
  is at most around a factor of 2 when 
  using a uniform grid with between 125 and 2,000 grid points.
  On the other hand, the densities are systematically higher 
  than those in fully resolved 1D models, which can be tracked 
  down to the simple model
  underestimating the true value of the integrated radiative 
  losses 
  in the UTR ($R_{utr}$), at the very start of the heating 
  phase. One can mitigate this problem by using slightly more 
  complicated models for $R_{utr}$ at the start of the 
  increased heating event and this will be addressed in a 
  subsequent publication. 
  However, for the present, the density 
  draining phase is captured correctly 
  which is important as 
  this is the phase that is seen in many observations of 
  coronal loops.
  We note that in Case 8, during this phase and
  throughout the entire evolution, the most refined
  uniform grid 
  solution (Lare1D with 8,000 grid points) achieved a better 
  agreement with the fully resolved model
  than the jump condition (LareJ with 500 grid points) 
  solution but at significantly greater computational cost.  
 \\
 \indent
 Our emphasis here has been on obtaining an improved coronal 
 density. This is important for interpreting observations of, 
 for example, active region loop cores, \lq warm\rq\ 
 loops, 
 as well as microflare and flare coronal emission. On the 
 other hand, by treating the lower (unresolved) TR as a 
 discontinuity, information will be lost on detailed TR 
 emission lines such as CIV. If the jump condition is applied 
 close to 1 MK (i.e. between $ 5 \times 10^5$ K and 1 MK) the 
 details of the (bright) TR will be lost, although integrated 
 TR quantities can of course still be deduced. This loss of 
 detail would particularly affect studies of, for example, the 
 bright TR “moss” – bright emission at the footpoints of very 
 hot loops 
  \citep[see e.g.][]{paper:Flectcher&DePontieu1999}.
  Full 
  numerical resolution is still required to deduce these, with 
  the corresponding risk of serious errors in the plasma 
  density. Model setups with smaller coronal domains (coronal 
  heights) and or lower temperatures (say below 1-2 MK) are 
  likely to have adequate resolution
  \citep[e.g.][]{paper:Zachariasetal2011,
  paper:Hansteenetal2015}.
  \\
  \indent
  In summary, this paper has presented an approach to deal 
  with the difficulty of obtaining the correct interaction 
  between a downward conductive flux from the corona and the 
  resulting upflow from the TR. A wide range of impulsive 
  (spatially uniform) heating events was considered for both 
  short and long loops. Our new method was used in simulations 
  with coarse resolutions that do not resolve the lower 
  transition region. The main result is that the method leads 
  to (i) coronal densities comparable to fully-resolved 1D 
  models but with significantly faster computation times, and 
  (ii) significant improvements in the accuracy of both the 
  coronal density and temperature temporal evolution when 
  compared to the equivalent simulations run without this 
  approach. 
  
%
%

\begin{acknowledgements} 
  The authors are grateful to Dr. Stephen Bradshaw for
  providing us with the HYDRAD code.
  We also thank the referee for their helpful 
  comments that improved the presentation.
  C.D.J. acknowledges the financial support of the Carnegie   
  Trust for the Universities of Scotland. 
  This project has received funding from the Science and 
  Technology Facilities Council (UK) through the consolidated 
  grant ST/N000609/1 and the European Research Council (ERC) 
  under the European Union's Horizon 2020 research and 
  innovation program (grant agreement No 647214).
\end{acknowledgements}
%
%
\begin{appendix}

%
%
%
\section{Lare1D with thermal conduction and radiation
  \label{app:A}}
  \indent
  The 1D field-aligned MHD equations 
  \eqref{eqn:1d_continuity}-\eqref{eqn:gas_law} are solved
  using a Lagrangian remap (Lare) approach, 
  as described for 3D MHD in
  \cite{paper:Arber2001}, adapted for 1D field-aligned
  hydrodynamics.
  Time-splitting methods 
  are used to split the field-aligned equations into an 
  ideal hyperbolic component and non-ideal components.
  This allows thermal conduction and optically 
  thin radiation to be updated separately from the advection 
  terms since these 
  effects formulate the non-ideal components.
  \\
  \indent
  During 
  a single time step, we first assume that we have no flows, 
  so 
  that only the temperature (specific-internal energy density) 
  can change, and update the temperature (specific-internal 
  energy density) based on the effects of thermal conduction, 
  optically 
  thin radiation and heating. 
  We then use a one-dimensional Lagrangian 
  remap method (Lare1D) to solve the field-aligned ideal MHD 
  equations, updating the pressure, density, velocity and 
  temperature (specific-internal energy density). 
  \\
  \indent
  The Lagrangian remap code 
  (Lare) splits each time step into a Lagrangian step followed 
  by a remap step. The Lagrangian step solves the ideal MHD 
  equations in a frame of reference that moves with the fluid.  
  By using time-splitting methods, thermal 
  conduction, optically thin radiation and heating have been
  included in the 
  Lagrangian step.  
  The remap step then maps the variables back onto the 
  original grid. 
%
%
%
%
\subsection{Field-aligned ideal MHD equations}
  \indent
  The Lare1D code solves the normalised field-aligned ideal 
  MHD 
  equations,
  \begin{align}
    & 
    \frac{\partial\rho}{\partial t} + v\frac{\partial\rho}
    {\partial z} = - \rho\frac{\partial v}{\partial z}, 
    \label{eqn:1d_continuity_Lare}
    \\[2mm]
    & 
    \rho \frac{\partial v}{\partial t} + \rho v
    \frac{\partial 
    v}
    {\partial z}
    = -
    \frac{\partial P}
    {\partial z} - \rho g_{\parallel}
    + \rho \nu \frac{\partial^2 v}{\partial z^2},
    \label{eqn:1d_motion_Lare}
    \\[2mm]
    & 
    \rho \frac{\partial\epsilon}{\partial t} + \rho v
    \frac{\partial
    \epsilon}
    {\partial z}
    = -P\frac{\partial 
    v}
    {\partial z}
    +    
    \rho \nu \!   
    \left(
    \frac{\partial v}{\partial z}
    \right)^{\!\!2},
    \label{eqn:1d_tee_Lare}
    \\[2mm]
    & 
    P = 2 \rho T,
    \label{eqn:gas_law_Lare}
  \end{align} 
  on a 
  staggered grid (velocities are defined at the cell 
  boundaries
  and all scalars are defined at the cell centres)
  using a predictor-corrector scheme that is second-order 
  accurate in both space 
  and time. This method stably integrates the solution, on an 
  advective time step that is governed by the 
  Courant-Friedrichs-Lewy  (CFL) condition, 
  \begin{align}
    \hspace{2cm}
    \Delta t_{adv} \leq 
    \dfrac{\Delta z} {\textrm{max}(\sqrt{c_s^2 + v^2})},
    \label{eqn:dt_adv}
  \end{align} 
  where $c_s$ is the local sound speed.

%
%
%
\subsection{Thermal conduction}
  \label{subsection:Thermal Conduction}
  \indent
  The thermal conduction model is based on the classical
  Spitzer-Harm heat flux formulation
  (\cite{paper:Spitzer1962}). In the time-splitting update,
  the thermal conduction step is of the form,
  \\
  \begin{align}
  \hspace{2cm}
    \rho \frac{\partial\epsilon}{\partial t}
    =
    -
    	\dfrac{\partial}{\partial z}
	\left( 
	-
	\kappa_0
	T^{5/2} \dfrac{\partial T}{\partial z}
	\right).
	\label{eqn:1d-tc}
  \end{align}  
  We treat 
  thermal conduction using the RKL2
  super time stepping (STS) method, as         
  described in \cite{paper:Meyeretal2012,paper:Meyeretal2014} 
  and discussed in
  Appendix \ref{app:B}.
  For the RKL2 method we
  approximate the parabolic conduction operator using central
  differencing of the heat flux,
  \begin{equation}
    \resizebox{.85\hsize}{!}{
    $
    {\bf{L}}^{\textsc{c}}(T)= 
    -
    \dfrac{1}{\rho}
    \dfrac{\partial}{\partial z}
    \left( 
    -
    \kappa_0
    T^{5/2} \dfrac{\partial T}{\partial z}
    \right) \approx
    -
    \dfrac{1}{\rho} \dfrac{F_{sp, \, i+ \frac{1}{2}} -
    F_{sp, \, i- \frac{1}{2}}}{dzb_i}
    $
    },
  \end{equation}
  where, 
  \begin{align}
    \hspace{0.5cm}
    F_{sp, \, i+ \frac{1}{2}} =
    -
    \kappa_0
    \left(
    \dfrac{T_{i+1}+T_{i}}{2}
    \right)^{5/2}
    \left(
    \dfrac{T_{i+1}-T_{i}}{dzc_i}
    \right),
  \end{align}
  and ${dzb_i}$ (${dzc_i}$) is the distance between cell 
  boundaries (centres).
  \\
  \indent
  The conductive flux-saturation limit describes the maximum
  heat flux that the plasma is capable of supporting
  \citep{paper:Bradshaw&Cargill2006}. This
  limit is reached when all of the particles travel in the
  same direction at the electron thermal speed, $v_{th}=(k_BT
  m_e)^{1/2}$, and is given by,
  \begin{align}
    \hspace{2cm}
    F_{sa}= \dfrac{3 }{2 m_p \sqrt{m_e}} 
    \rho (k_{B}T)^{3/2},
  \end{align}
  where $m_p$ and $m_e$ are the proton and electron masses,
  respectively.
  In our numerical simulations, heat flux limiting is 
  important because there is a sufficient amount 
  of 
  heating, in many of the events considered, so that the 
  Spitzer-Harm
  heat flux, 
  \begin{align}
    \hspace{2cm}
    F_{sp}= - \kappa_0 T^{5/2}
    \frac{\partial T}{ \partial z},
  \end{align}
  can exceed the conductive flux-saturation limit.
  Therefore, we impose the following heat flux limiter that 
  was 
  described 
  in
  BC13, 
  \begin{align}
    \hspace{2cm}
    F_{c}= \dfrac{\quad F_{sp} \times F_{sa}}
    {\sqrt{F_{sp}^2 + F_{sa}^2}},
  \end{align}
  to limit the Spitzer-Harm heat flux.
  \begin{figure}
    \centering
    \resizebox{\hsize}{!}
    {\includegraphics
    {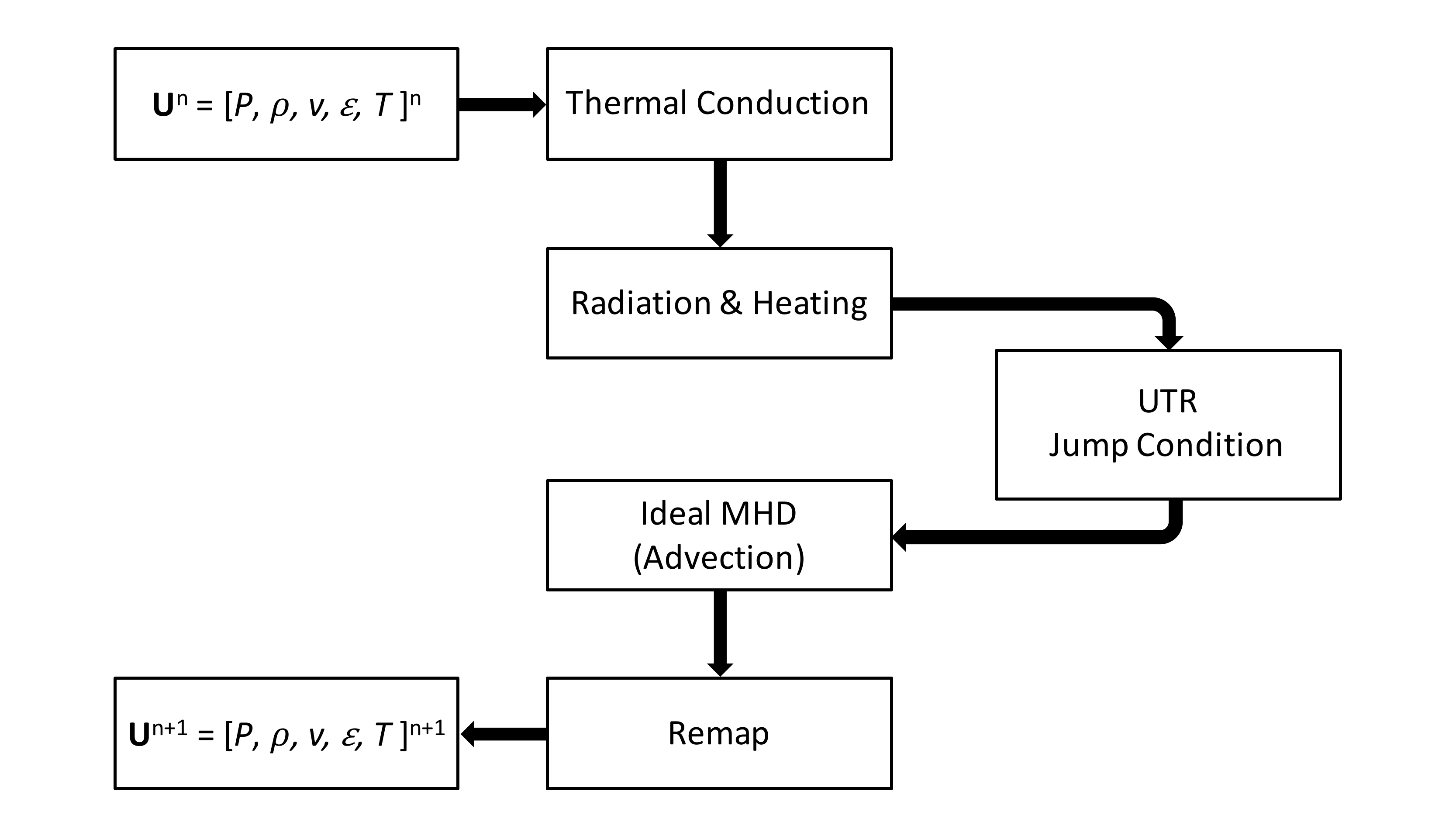}}
    \caption{
    Lare1D with Thermal Conduction and Radiation
    time-splitting update strategy. The modification to
    include
    the UTR jump condition is also outlined. This step
    is ignored when the
    Lare1D code is employed without the jump condition.
    \label{Fig:Lare_update}
    }
  \end{figure}
  %
  %
  
%
%
%
%
\subsection{Optically thin radiation (OTR)}
  \label{subsection:OTR}
  \indent
  For the optically thin radiative loss function we use a 
  piecewise continuous power law,
  \begin{align}
    \hspace{3cm}
    L_r=n^2\chi T^\alpha,
  \end{align}
  where the
  temperature dependent constants $\chi$ and $\alpha$
  are defined
  following \cite{paper:Klimchuketal2008}.
  In the time-splitting update, the radiation step is of the 
  form,
  \begin{align}
    \hspace{3cm}
    \rho \frac{\partial\epsilon}{\partial t}
    = - 
    n^2\chi T^\alpha,
    \label{eqn:1d-r}
  \end{align} 
  which is integrated using a time-centred finite difference 
  method (FDM). 
  To prevent the plasma from catastrophically cooling 
  under
  the effects of OTR, we impose a radiative time step
  restriction, $\Delta t_{rad}$, on the integration,
   that prevents the temperature
  (specific internal energy density) from decreasing by more
  than $1\%$
  during a single time step.
  This radiative restriction is not as severe as the advective
  time step \eqref{eqn:dt_adv} but can become important at the 
  peak of the radiative losses.
  \\
  \indent
  To maintain our isothermal chromosphere, at a 
  temperature of 10,000K, radiation is smoothly turned off 
  over a $100$K interval,
  above
  the chromospheric temperature
  \citep[][BC13]{paper:Klimchuketal1987}.
%
%
%
%
\subsection{Heating}
  \indent
  The Lare code deals with the effects of viscous heating 
  during the advection step. However, we also include a 
  separate 
  heating step of the form,
  \begin{align}
    \hspace{3cm}
    \rho \frac{\partial\epsilon}{\partial t}
    = Q,
    \label{eqn:1d-h}
  \end{align}  
  where our heating function, which is the dominant source of 
  heating 
  in our numerical simulations, is defined as the sum 
  of contributions from both the 
  background heating ($Q_{bg}$) and additional heating 
  ($Q_{H}$),
  \begin{align}
    \hspace{3cm}
    Q=Q_{bg}+Q_{H}.
  \end{align} 
  The heating step is integrated using a simple FDM which we
  incorporate into the radiation step \eqref{eqn:1d-r}. This 
  allows 
  the temperature (specific internal energy
  density) to be updated due to the effects of optically thin 
  radiation and
  heating simultaneously.

%
%
%
%
\subsection{Time-splitting update}
  \indent
  Let ${\bf U}=[P,\rho,v,\epsilon, T]$, be a vector of the   
  model variables. The one-dimensional field-aligned MHD   
  equations can then be written in terms of an ideal MHD 
  component and non-ideal   
  components,
  \begin{align}
    \hspace{1cm}
    \dfrac{\partial {\bf U}}{\partial t} = 
    {\bf L}^{\textsc{c}}
    ({\bf U}) + {\bf L}^{\textsc{r}}({\bf U}) + 
    {\bf L}^{\textsc{mhd}}({\bf U}),
  \label{eqn:1d-ts}
  \end{align}
  where ${\bf L}^{\textsc{c}}$, ${\bf L}^{\textsc{r}}$
  and ${\bf L}^{\textsc{mhd}}$ are the thermal conduction, 
  radiation and heating and ideal MHD operators respectively.
  During a single time step, we use the Lie-splitting 
  (sequential splitting) method \citep{paper:Faragoetal2011} 
  to integrate these operators separately.
  \\
  \indent
  The
  temperature (specific internal energy
  density) is updated first,
  based on the effects of thermal conduction, 
  OTR 
  and heating, before the ideal field-aligned MHD 
  equations are solved. Following this strategy, 
  the Lie-splitting update for one complete time step is given
  by,
  \begin{eqnarray}
    \hspace{2cm}
    \bf{U}^* &=& {\bf{ C(U}}^n, \Delta  t),
    \nonumber
    \\[2mm]
    \bf{U}^{**} &=& {\bf{ R(U}}^*,\Delta  t),
    \nonumber
    \\[2mm]
    \bf{U}^{n+1} &=& {\bf{ MHD(U}}^{**},\Delta  t),
    \label{eqn:1d-tsu}
  \end{eqnarray}
  where ${\bf{U}}^{n+1}={\bf{C}(U}^n,\Delta t)$,
  ${\bf{U}}^{n+1}={\bf{R}(U}^n,\Delta t)$ 
  and ${\bf{U}}^{n+1}={\bf{MHD}(U}^n,\Delta t)$
  represent the 
  updates of thermal conduction, radiation and heating and
  ideal MHD,
  for the time step $\Delta t$. This update strategy is shown
  in Fig. \ref{Fig:Lare_update}
  \\
  \indent
  Since we treat thermal conduction using STS methods 
  we super-step the conductive timescale
  restriction (accelerate the explicit sub-cycling).
  Therefore, 
  the time-splitting strategy 
  \eqref{eqn:1d-tsu} stably integrates the 
  field-aligned MHD equations, 
  on a time step that is given by,
  \begin{align}
    \hspace{2cm}
    \Delta t = \textrm{min} \, (\Delta t_{adv},\, 
    \Delta t_{rad}).
  \end{align}
  %
  %

%
%
%
%
\section{Super time stepping methods to treat thermal   conduction
  \label{app:B}}
  \begin{table}
    \caption{
    \label{table:sts_ct}
    Numerical simulation computation times 
    (run on a single processor)
    for three
    different methods to treat thermal conduction.
    }
    \centering
    \resizebox{\hsize}{!}
    {
    \begin{tabular}{lcccccc}
    \hline\hline
    Case & $N_z$
    & $\tau_{\textrm{sts}}$
    & $\tau_{\textrm{cyc}}$
    & $\tau_{\textrm{exp}}$
    & $\tau_{\textrm{cyc}}${\tiny /}
    & $\tau_{\textrm{exp}}${\tiny /}
    \\
    & &
    (mins) & (mins) & (mins) 
    & $\tau_{\textrm{sts}}$
    & $\tau_{\textrm{sts}}$
    \\
    \hline
    1  & 500   &2.45  &1.98    &2.25  & 0.81 & 0.92
    \\ 
       & 1,000 &6.73  &6.47    &15.72 & 0.96 & 2.34
    \\
       & 2,000 &12.23 &29.07   &128   & 2.38 & 10.5
    \\
       & 4,000 &42.6  &199     &592   & 4.67 & 13.9
    \\
       & 8,000 &205   &1,537   &4,699 & 7.50 & 22.9
    \\
    \hline
    2  & 500   &6.32  &8.12    &25.7  & 1.28 & 4.07
    \\ 
       & 1,000 &18.5  &45.02   &122   & 2.43 & 6.59
    \\
       & 2,000 &48.8  &308     &970   & 6.31 & 19.9
    \\
       & 4,000 &135   &2,385   &7,772 & 17.7 & 57.6
    \\
       & 8,000 &607   &18,778  &47,123*& 30.9 & 77.6
    \\
    \hline
    3  & 500   &12.15 &33.13   &168   & 2.73 & 13.8
    \\ 
       & 1,000 &49.67 &257     &790   & 5.17 & 15.9
    \\
       & 2,000 &138   &2,023   &6,238 & 14.7 & 45.2
    \\
       & 4,000 &579   &15,958  &48,405*& 27.6 & 83.6
    \\
       & 8,000 &2,440 &108,898*&238,620*&44.6 & 97.8
    \\
    \hline
    \end{tabular}
    }
    \tablefoot{
    The columns show
    the number of grid points (uniform grid used), 
    the computation times by treating thermal conduction using
    super time stepping methods (sts), 
    explicit time step sub-cycling (cyc) and
    explicit time step evolution (exp), 
    and the computation
    time ratios between these methods.
    The simulations (Cases 1-3 of Table
    \ref{table:simulations}) are run to a final 
    time of 60s, which coincides with the end of the heating 
    period.
    The asterisks indicate runs where the computation time
    to the final time has been estimated based on results over 
    a shorter period.   
    }
  \end{table}
  \begin{figure}
    \hspace{-0.5 cm}
    \resizebox{\hsize}{!}
    {\includegraphics{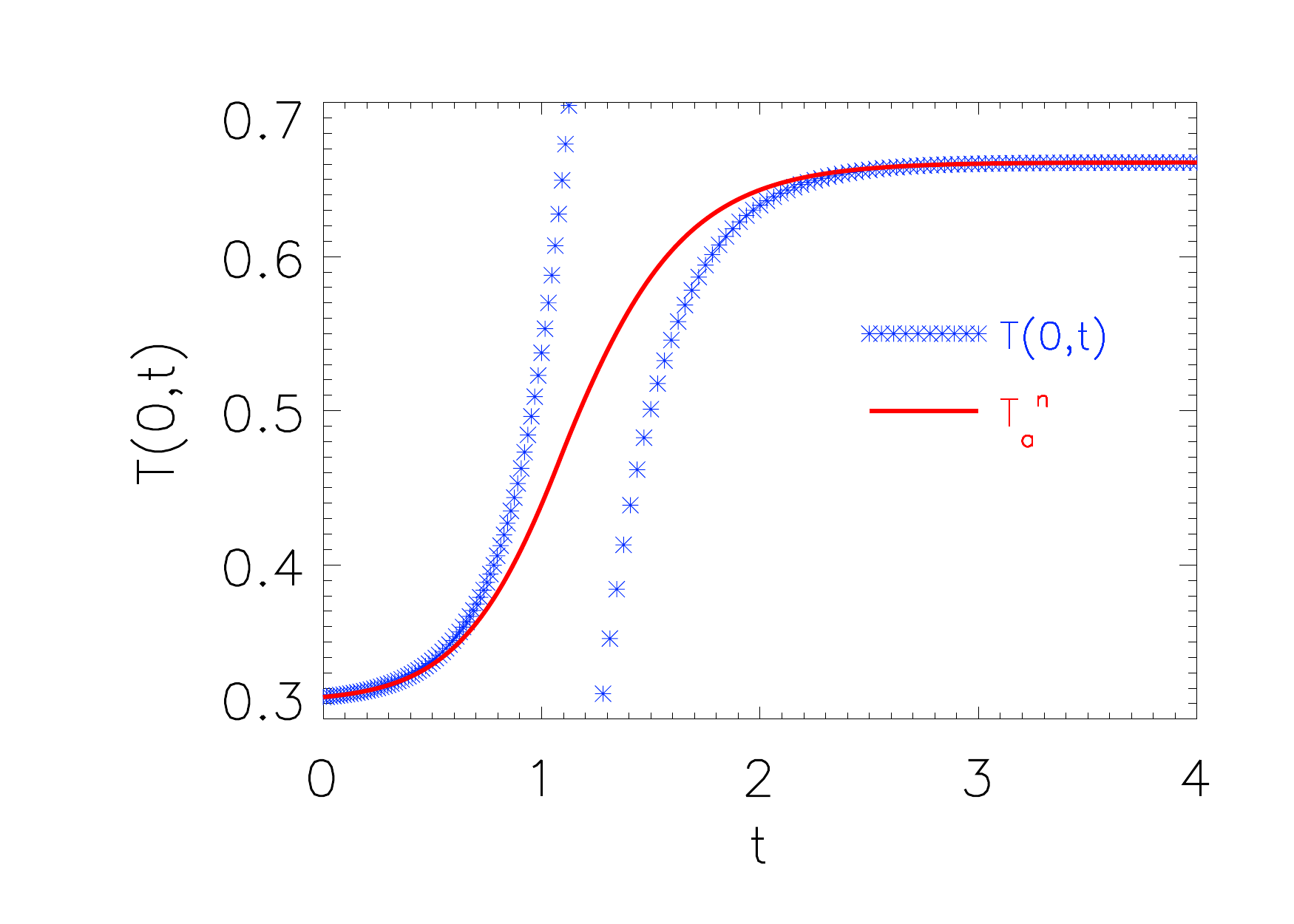}}
    \vspace{-0.5cm}
    \caption{
    \label{Fig:thermal_ins}
    The panel shows the 
    temporal evolution of $T(0,t)$.
    The solution
    leaves a thermally unstable
    isothermal equilibrium 
    and approaches a new 
    stable, non-isothermal
    equilibrium.
    The solid red curve is the numerical solution
    obtained by using the RKL2 STS method ($T^n_{a}$)
    and the blue asterisks represent
    the corresponding linear solutions ($T(0,t)$).
    The units on both axes are arbitrary.
    }
  \end{figure}
  \indent
  In the interests of computational efficiency,
  to relax the conductive timescale stability restriction of
  an explicit method 
  $
  (
  \Delta t_{cond} \leq
  \rho (\Delta z)^2 
  /
  (2\kappa_0
  T^{5/2})
  )
  $, we treat 
  thermal conduction by using
  super time stepping (STS) methods, as         
  described in 
  \cite{paper:Meyeretal2012,paper:Meyeretal2014}. 
  These methods are essentially an acceleration
  of explicit time step sub-cycling and have been used 
  effectively to speed up the integration of parabolic 
  operators. 
  In particularly, we use the 
  Runge-Kutta Legendre method with second-order temporal
  accuracy (RKL2). 
  \\
  \indent 
  Extending on the test problems considered in 
  \cite{paper:Meyeretal2012,paper:Meyeretal2014},    
  we have tested the RKL2 method for
  appropriateness of  use in coronal plasma 
  conditions, in order to ensure that the increased
  conductive time step does not 
  influence the correct temporal evolution.
  The Zel'dovich problem of a propagating 
  conduction front \citep{paper:Zeldovich1967} has been 
  solved.
  \\
  \indent
  In addition,
  we investigate whether or not STS methods can correctly 
  obtain 
  the
  growth (decay) rate when leaving (approaching) a thermally
  unstable (stable) isothermal (non-isothermal) equilibrium.
  Using a model equation, under the assumption of constant
  density, we solve the boundary value problem,
  \begin{align}
  &
  \dfrac{\partial T}{\partial t}
  =
  -
  \dfrac{\partial}{\partial z}
  \left(
  -
  T^{5/2} \dfrac{\partial T}{\partial z}
  \right) -\chi T^\alpha + H, \ -1/2\leq z \leq 1/2,
  \nonumber
  \\[2mm]
  &
  T(-1/2,t) = T(1/2,t) = T_0,
  \label{eq:nlcrh}
  \end{align}
  with the initial condition,
  \begin{align}
  T(z,0) = T_0 + \bar{T_1} \cos(\pi z),
  \quad -1/2\leq z \leq 1/2.
  \nonumber
  \end{align}
  $T_0$ is the isothermal unstable equilibrium
  and $\bar{T_1} \cos(\pi z)$ is a small perturbation.
  Linearising equation \eqref{eq:nlcrh},
  the temperature grows as,
  \begin{align}
  &
  \hspace{2cm}
  T(0,t) = T_0 + \bar{T_1}(0)e^{\sigma t},
  \label{eq:nlcrh_gr}
  \end{align}
  with $\sigma = -\pi^2 T_0^{5/2} 
  - \alpha {\chi} T_0^{{\alpha} - 1}$.
  Fig. \ref{Fig:thermal_ins} shows the temporal evolution of 
  $T(0,t)$ using the STS method, as a solid red curve labelled
  $T_a^n$. 
  The linear solution \eqref{eq:nlcrh_gr} is shown as 
  asterisks
  and the exact growth rate matches the rate calculated from
  the computational solution. A similar analysis confirms that 
  the exact decay rate, as the temperature evolves towards
  the non-isothermal stable equilibrium, is also correctly 
  predicted by the STS method.
  Therefore, we believe that 
  STS methods are appropriate 
  for use in solving more complex coronal plasma based 
  problems, 
  where the effect of thermal conduction plays an 
  important role. 
  \\
  \indent
  Although STS methods have already been 
  implemented in some 3D MHD codes
  \citep[e.g.][in press.]{paper:Realeetal2016}
  it remains instructive here to present a
  quantification of the computational gains involved.
  Based on the computation time ratios in Table
  \ref{table:sts_ct}, the benefit of using STS methods is 
  immediately clear, especially as the coronal temperature,
  which scales 
  strongly with the heating event, increases and the  
  conductive timescale 
  decreases.

\end{appendix}

%
%
\bibliographystyle{aa}
\bibliography{LareJPaper1}

\end{document}